\documentclass[aps,prb,floatfix,amsmath,amssymb,preprint,eqsecnum,nofootinbib,superscriptaddress]{revtex4}
\usepackage{graphicx}
\usepackage{dcolumn}
\usepackage{bm}
\usepackage{epstopdf}
\usepackage{setspace}   

\newcommand{\beq}{\begin{equation}}
\newcommand{\eeq}{\end{equation}}
\newcommand{\ba}{\begin{array}{ccc}}
\newcommand{\ea}{\end{array}}
\newcommand{\nn}{\nonumber}
 \renewcommand{\d}{\partial}
\def\bea{\begin{eqnarray}}
\def\eea{\end{eqnarray}}

\def\prl{\parallel}

\def\<{\langle}
\def\>{\rangle}
\usepackage{amsmath}
\usepackage{amssymb}
\begin{document}

\title{Entanglement Entropy in the $O(N)$ model}

\author{Max A. Metlitski}
\affiliation{Department of Physics, Harvard University, Cambridge MA
02138}

\author{Carlos A. Fuertes}
\affiliation{Instituto de F\'\i sica Te\'orica IFT UAM/CSIC,
Facultad de Ciencias C-XVI, C.U. Cantoblanco, E-28049 Madrid, Spain}

\author{Subir Sachdev}
\affiliation{Department of Physics, Harvard University, Cambridge MA
02138}

\date{\today\\
\vspace{1.6in}}
\begin{abstract}
It is generally believed that in spatial dimension $d > 1$ the leading contribution to the entanglement entropy $S = - tr \rho_A \log \rho_A$ scales as the area of the boundary of subsystem $A$. The coefficient of this ``area law" is non-universal. However, in the neighbourhood of a quantum critical point $S$ is believed to possess subleading universal corrections. In the present work, we study the entanglement entropy in the quantum $O(N)$ model in 
$1 < d < 3$. We use an expansion in $\epsilon = 3-d$ to evaluate ({\em i\/}) the universal geometric correction to $S$ for an infinite cylinder divided along a circular boundary; ({\em ii\/}) the universal correction to $S$ due to a finite correlation length. Both corrections are different at the Wilson-Fisher and Gaussian fixed points, and the $\epsilon \rightarrow 0$ limit of the Wilson-Fisher fixed point is distinct from the Gaussian fixed point.
In addition, we compute the correlation length correction to the Renyi entropy $S_n = \frac{1}{1-n} \log tr \rho^n_A$ in  $\epsilon$ and large-$N$ expansions. For $N \to \infty$, this correction generally scales as $N^2$ rather than the naively expected $N$. Moreover, the Renyi entropy has a phase transition as a function of $n$ for $d$ close to 3.

\end{abstract}

\maketitle

\section{Introduction}
One of the most fascinating and counterintuitive properties of a quantum system is the entanglement of its many-body wave-function. In recent years, there has been a lot of interest in using entanglement as a theoretical probe of ground state correlations. It is hoped that this viewpoint will be particularly fruitful in studying quantum critical points, which realize some of the most non-classical, entangled states of matter.

A useful measure of entanglement is given by the entanglement entropy $S$, also known as von-Neumann entropy. To compute $S$, we divide the system into two parts, $A$ and $B$, and determine the reduced density matrix $\rho_A = tr_B \rho$, where $\rho$ is the full density matrix of the system. Then, the entanglement entropy,
\beq S_A = - tr_A \rho_A \log \rho_A \eeq
If the system is in a pure state, then the entanglement entropy is ``mutual", i.e. $S_A = S_B$. 

One may ask how does the entanglement entropy behave near a quantum critical point. This question has been addressed completely for one-dimensional critical points with dynamical critical exponent $z = 1$. Such critical points are described by $1+1$ dimensional conformal field theories (CFT's). In these systems if $A$ is chosen to be a segment of length $l$ and $B$ - its complement in the real line, the entanglement entropy is given by,\cite{Holzhey,Cardy}
\beq S = \frac{c}{3} \log l/a \label{Sc}\eeq
where $a$ is the short-distance cut-off and the constant $c$, known as the central charge, is a fundamental property of the CFT. 
Moreover, if the system is perturbed away from the critical point, the entanglement entropy becomes,
\beq S = {\cal A} \frac{c}{6} \log \xi/a \label{Sxi}\eeq
where $\xi$ is the correlation length and ${\cal A}$ is the number of boundary points of the region $A$. Here it is assumed that $A$ and $B$ are composed of intervals whose length is much larger than $\xi$. 

The study of entanglement entropy at quantum critical points in dimension $d > 1$ has received much less attention. The leading contribution to $S$ is believed to satisfy the ``area law",\cite{Srednicki}
\beq S = C  \frac{{\cal A}}{a^{d-1}} \label{SA}\eeq
where ${\cal A}$ is the length/area of the boundary between the regions $A$ and $B$. Physically, the area law implies that the entanglement in $d > 1$ is local to the boundary even at the critical point. The coefficient $C$ entering the area law is sensitive to the short distance cut-off, and is, therefore, non-universal. So, in contrast to the one-dimensional case, the leading term (\ref{SA}) in the entanglement entropy in higher dimensions cannot be used to  characterize various critical points.

However, one may subtract the leading non-universal area-law contribution to the entanglement entropy and consider,
\beq \Delta S = S - C \frac{{\cal A}}{a^{d-1}}\eeq
At least for Lorentz-invariant theories that we study here, it is generally believed that if $d=2$ and the boundary between the regions $A$ and $B$ is closed and smooth, $\Delta S$ is universal. (Additional logarithmic divergences are believed to occur when the boundary contains corners/endpoints.\cite{Casini,Fursaev}) In particular, precisely at the critical point, $\Delta S$ is just a geometric constant. Moreover, $\Delta S$ is expected to remain universal when the theory is perturbed away from the critical point by a finite correlation length $\xi$. 

We note that the above considerations have only been verified by explicit field theoretic calculations in free theories. These assertions were also confirmed in strongly coupled supersymmetric gauge theories using the AdS/CFT correspondence.\cite{Ryu,Ryu2}
Recently, Hsu {\em et al.} \cite{Fradkin} found 
universal corrections for  a special class of quantum critical points in $d=2$
which are described by dimensional reduction to a classical $d=2$ field theory. However, such critical points
are non-generic, and unstable \cite{senthil} in physical situations to quantum critical points described by
interacting field theories in 3 space-time dimensions. 

In the present work, we compute the geometric and correlation length corrections to the entanglement entropy in the simplest generic interacting CFT in $d = 2$ dimensions - the $O(N)$ model. We verify that these corrections are, indeed, universal. We perform our calculations using expansions in $\epsilon = 3-d$ and $1/N$. Note that the universality of $\Delta S$ formally extends to the range $2 < D < 4$, where $D = d + 1$ is the space-time dimension.\footnote{In $D = 4$, $S$ develops new singularities associated with the extrinsic curvature of the boundary.\cite{Ryu2}}

\begin{figure}[t]
\begin{center}
\includegraphics[width=4in]{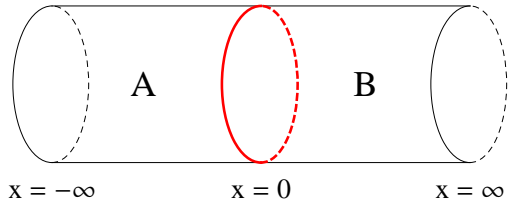}
\caption{The cylindrical geometry considered in calculation of finite size correction to the entanglement entropy.}
\label{Figgeom}
\end{center}
\end{figure}
In the rest of this paper we consider the following geometry. We take two semi-infinite regions $A$ and $B$ with a straight boundary at $\mbox{x} = 0$. The boundary extends along the remaining $d - 1$ spatial directions, each taken to have a length $L$. For technical reasons, we impose anti-periodic boundary conditions along each of these directions. We also consider more general boundary conditions with a twist by an arbitrary phase $\varphi$ in a theory of $N/2$ complex scalar fields. So in the physical case $d = 2$, our space is an infinite cylinder divided into regions $A$ and $B$ along a circle of length $L$, see Fig. \ref{Figgeom}. In this geometry the entanglement entropy at the critical point is given by,
\beq S = C \frac{L^{d-1}}{a^{d-1}} + \gamma \label{gammadef}\eeq
We explicitly compute the universal geometric constant $\gamma$. To leading order in $4-\epsilon$ expansion we obtain,
\beq \gamma = - \frac{N \epsilon}{6 (N+8)} \left( \log\Big|\theta_1\big(\frac{\varphi(1+i)}{2 \pi}, i\big)\Big| - \frac{\varphi^2}{4 \pi} - \log \eta(i)\right), \quad D = 4-\epsilon, \mbox{ Wilson-Fisher fixed point}\label{gammaval} \eeq
Here $\theta_1$ and $\eta$ are Jacobi elliptic and Dedekind-eta functions. The sign of $\gamma$ depends on the value of $\varphi$: it is negative for $\varphi = \pi$ (anti-periodic boundary conditions) and positive for $\varphi \to 0$. Note that eq. (\ref{gammaval}) is only valid for $\varphi \gg \epsilon^{1/2}$. For zero twist (periodic boundary conditions), we hypothesize that to leading order,
\beq \gamma = -\frac{N \epsilon}{12 (N+8)} \log \epsilon \label{gammaloge} \eeq
The result (\ref{gammaval}) should be compared to the corresponding value at the Gaussian fixed point in $4-\epsilon$ dimensions,
\beq \gamma = -\frac{N}{6} \left( \log\Big|\theta_1\big(\frac{\varphi (1+i)}{2 \pi}, i\big)\Big| - \frac{\varphi^2}{4 \pi} - \log \eta(i)\right), \quad D = 4-\epsilon, \mbox{ Gaussian fixed point}\eeq
We see that $|\gamma|$ is parametrically smaller at the Wilson-Fisher fixed point than at the Gaussian fixed point. 
Thus, entanglement entropy distinguishes these two fixed points already at leading order in $\epsilon$ expansion. 

If we perturb the system away from the critical point, we can take the limit $L \to \infty$ and obtain the general scaling relation,
\beq S = C \frac{L^{d-1}}{a^{d-1}} + r \frac{L^{d-1}}{\xi^{d-1}}\label{Sxid}\eeq
where $r$ is a universal coefficient that we compute. In general, one has to make a specific choice for the definition of the correlation length $\xi$. In the $O(N)$ model there is a very natural choice, $\xi = m^{-1}$, where $m$ is the gap to the first excitation. Note that in the present work we only consider the phase of the $O(N)$ model with unbroken symmetry. The value of $r$ to leading order in $4-\epsilon$ expansion is found to be,
\beq r = - \frac{N}{144 \pi}, \quad D = 4-\epsilon, \mbox{ Wilson-Fisher fixed point}\eeq
As with the finite size correction, $r$ is parametrically smaller at the Wilson-Fisher fixed point than at the Gaussian fixed point where,\cite{Cardy}
\beq r = - \frac{N}{24 \pi \epsilon},\quad D = 4-\epsilon, \mbox{ Gaussian fixed point} \eeq

In addition to the entanglement entropy, we study the Renyi entropy,
\beq S_n = \frac{1}{1-n} \log tr_A \rho_A^n\eeq
The Renyi entropy always naturally appears in  field-theoretic calculations as it is related to the partition function of the theory on an $n$-sheeted Riemann surface. 
One then obtains the entanglement entropy by taking the limit, $S = \lim_{n \to 1} S_n$. At least for $n$ close to $1$, the Renyi entropy is believed to possess the same universal properties as the entanglement entropy. In particular, the finite size and correlation length corrections are given by,
\bea 
S_n &=& C_n \frac{L^{d-1}}{a^{d-1}} + \gamma_n \label{gammandef}\\
S_n &=& C_n \frac{L^{d-1}}{a^{d-1}} + r_n \frac{L^{d-1}}{\xi^{d-1}}\label{sndef}\eea
where the non-universal coefficient $C_n$ of the leading area law term, as well as the universal coefficients $\gamma_n$, $r_n$ are now $n$ dependent. We compute $r_n$ in $4-\epsilon$ and large-$N$ expansions. A careful renormalization group analysis demonstrates that $r_n$ is parametrically enhanced in both of these limits. In particular, $r_n \sim O(\frac{1}{\epsilon})$ in the $4-\epsilon$ expansion. However, the enhancement is most striking in the large-$N$ expansion where we find $r_n \sim O(N^2)$. Such scaling is in contrast with the result $r_n \sim O(N)$ that one would obtain at each order in $1/N$ for fixed correlation length $\xi$, implying that the limits $\xi \to \infty$ and $N \to \infty$ do not commute. As far as we know, this is the first violation of naive large-$N$ counting in the $O(N)$ model. A common feature of the two expansions is that the leading term of $r_n$ behaves as $r_n \sim n - 1$ for $n \to 1$ and does not contribute to the entanglement entropy $S$. Hence, $r \sim O(N)$ in the large $N$ limit and $r \sim O(1)$ in the $4 - \epsilon$ expansion. 

Another unusual phenomenon that we find in $4 -\epsilon$ expansion is non-analytic dependence of the coefficients $\gamma_n$, $r_n$ on $n$. In fact, $\gamma_n$ and $r_n$ will have a discontinuity at $n = n^*$, where $n^*$ is generally non-universal and lies in the range, $1 < n^* \le 1 + \frac{3}{4} \frac{N+2}{N+8} \epsilon$. The $n$-dependence of $\gamma_n$ and $r_n$ for $n < n^*$ and $n > n^*$ is, however, universal. Thus, we have two universal branches for $\gamma_n$ and $r_n$. We note that eqs. (\ref{gammandef}) and (\ref{sndef}) are understood in the limit $L \to \infty$, $\xi \to \infty$. However, there appears a new divergent length-scale in the problem as $n \to n^*$, and the limits $n \to n^*$ and $L \to \infty$, $\xi \to \infty$ do not commute. In particular, if we fix the size of our regions $L$ or the correlation length $\xi$, the $n$-dependence of the Renyi entropy $S_n$ will be completely analytic. Moreover, due to the emergence of a new length-scale as $n \to n^*$, in the crossover region $S_n$  is not entirely universal. We stress that any non-analyticity and non-universality only occurs away from the point $n = 1$. In particular, the entanglement entropy $S= \lim_{n \to 1} S_n$ is well defined and universal.

The non-analytic behaviour discussed above is also found to occur in the large-$N$ expansion in dimensions $2.74 \lesssim d < 3$. The limited range of $d$ suggests that this phenomenon might be absent in the $O(N)$ model in the physically relevant case $d = 2$. Nevertheless, we expect that such non-trivial $n$ dependence will occur quite generically at other quantum critical points.

This paper is organized as follows. In section \ref{sec:replica}, we remind the reader of the replica trick, which relates the entanglement entropy to the partition function on an $n$-sheeted Riemann surface. In section \ref{sec:general}, we show that the coefficient of the correlation length correction to the Renyi entropy $r_n$ is parametrically enhanced in both expansions we consider. Sections \ref{sec:4eps} and \ref{sec:4epsL} are respectively devoted to the evaluation of correlation length and finite size corrections in $4-\epsilon$ expansion. In section \ref{sec:largeN} we compute the coefficient $r_n$ in the large-$N$ expansion. Some concluding remarks are given in section \ref{sec:concl}.



\section{The replica trick}\label{sec:replica}
We consider the $O(N)$ model in $D = d+1$ space-time dimensions. The action for the $N$-component real scalar field $\phi$ is given by ,
\beq S = \int d^dx d \tau \left(\frac{1}{2} (\d_{\mu} \phi)^2 + \frac{t}{2} \phi^2 + \frac{u}{4} \phi^4\right)\label{soft}\eeq
We divide our space into two regions $A$ and $B$ with the boundary being a $d-1$ dimensional plane at $\mbox{x} = 0$. We will denote the coordinates along the boundary directions by $x_{\perp}$. The Renyi entropy $S_n$ may be calculated as,
\beq S_n = \frac{1}{1-n} \log \frac{Z_n}{Z_1^n} \label{diffn1}\eeq
from which we obtain the entanglement entropy,
\beq S = \lim_{n\to 1} S_n\label{SSn}\eeq 
Here $Z_n$ is the partition function of the theory on an $n$-sheeted Riemann surface. This Riemann surface lies in the $x_{\prl} = (\tau, \mbox{x})$ plane and has a conical singularity at $(\tau, \mbox{x}) = (0,0)$. The surface is invariant under translations along the $x_{\perp}$ directions. We may use the following metric for our space-time,
\beq ds^2 = dr^2 + r^2 d\theta^2 + dx^2_{\perp}\label{metric}\eeq
where $r, \theta$ are the polar coordinates in the $(\tau, \mbox{x})$ plane. Concentrating on this plane, we see that the metric is exactly the same as for the usual Euclidean plane; the only modification is that the angular variable $\theta$ has a period $\theta \sim \theta + 2 \pi n$.

\section{Parametric Enhancement of Correlation Length Correction}\label{sec:general}
In this section, we show that the coefficient $r_n$ of the correlation length correction to the Renyi entropy, eq. (\ref{sndef}), is parametrically enhanced in both expansions that we consider. Moreover, we demonstrate that $r_n$ can to leading order be extracted from the properties of the theory at the critical point. 

We start with the $O(N)$ model perturbed away from the critical point $t = t_c$ by a finite $\tilde{t} = t - t_c > 0$ (we drop the tilde below). To compute $r_n$, we need to find the dependence of the partition function $Z_n$ on the mass gap $m = \xi^{-1}$. Here we assume that the dimensions of the boundary $L \gg \xi$, so that we can take the limit $L \to \infty$. It is useful to differentiate,
\bea \frac{d}{dt} \log{\frac{Z_n}{Z_1^n}} &=& - \frac{1}{2} \left( \int_{n-\mathrm{sheets}} d^D x \, \langle \phi^2(x) \rangle_n - n \int_{1-\mathrm{sheet}} d^D x \,\langle \phi^2(x) \rangle_1\right)\\ 
&=& - \frac{1}{2} L^{d-1} \int_{n-\mathrm{sheets}} d^2x_{\prl}\, (\langle \phi^2(x) \rangle_n -   \langle \phi^2(x) \rangle_1) \label{ddt}\eea
where we have used the fact that the contribution to the integral from each of the sheets is the same (from here on, all integrals over $d^2 x_{\prl}$ are understood to be over $n$-sheets). Now, recalling, $m \sim t^{\nu}$, we may convert the derivative with respect to $t$ into a derivative with respect to $m$,
\beq m \frac{d}{dm}  \log{\frac{Z_n}{Z_1^n}} = - \frac{1}{2 \nu} L^{d-1} \int d^2 x_{\prl} \, t(\langle \phi^2(x) \rangle_n -   \langle \phi^2(x) \rangle_1) \label{mdm}\eeq
The expression $t(\langle \phi^2(x) \rangle_n -   \langle \phi^2(x) \rangle_1)$ is renormalization group invariant.\footnote{Two subtractions (constant and linear in $t$), in addition to the multiplicative renormalization, are needed to render the operator $\phi^2$ finite. However, these subtractions cancel among the two expectation values in (\ref{mdm}).} Thus, we may write,
\beq t(\langle \phi^2(x) \rangle_n -   \langle \phi^2(x) \rangle_1) = m^{D} f_n(m r)\label{tphi2}\eeq
where $f_n$ is a universal function. The function $f_n$ is expected to decay exponentially
for $m r \gg 1$, and the integral in (\ref{mdm}) converges for $r \to \infty$. The short-distance asymptotic of $f_n$ is controlled by the critical point. From the scaling dimension of the operator $\phi^2(x)$, $[\phi^2(x)] = D - \nu^{-1}$, we conclude,
\beq f_n(u) \to \frac{d_n}{u^{D-1/\nu}}, \quad u \ll 1 \label{fshort}\eeq
where $d_n$ is a universal constant. So the integral in (\ref{mdm}) converges for $r \to 0$, provided that $\nu^{-1} > D - 2$.\footnote{Otherwise, a $UV$ divergence appears which adds a piece analytic in $t$ to the entanglement entropy, in addition to the singular contributions discussed below.} In the $O(N)$ model in both expansions we consider, $\nu^{-1} = D -2 +  \nu_1$, where the correction $\nu_1$ is given to leading order by,
\bea
\nu_1 &=& \frac{6 \epsilon}{N+8}, \quad \quad D = 4-\epsilon \label{nu14eps}\\
\nu_1 &=& \frac{1}{N}\frac{8 \Gamma(D)}{D \Gamma(2-D/2) \Gamma(D/2-1)^2 \Gamma(D/2)}, \quad \nu_1(D =3) = \frac{32}{3 \pi^2 N}, \quad N \to \infty\eea
In particular, $\nu_1 > 0$ and $\nu^{-1}$ asymptotically approaches $D-2$ from above in both limits. 
With these remarks in mind, we integrate eq. (\ref{mdm}) with respect to $m$,
\beq \log{\frac{Z_n}{Z_1^n}}(t) - \log{\frac{Z_n}{Z_1^n}}(t=0) = - \frac{\pi n}{\nu (d-1)} (m L)^{d-1} \int_0^{\infty} du \, u f_n(u) \label{ffinal}\eeq
This is as far as we can proceed in general - to make further progress one needs the function $f_n(u)$. However, we have already noted that due to the fact, $\nu^{-1} \to D-2$, the integral in (\ref{ffinal}) is very close to diverging in both expansions. Hence, to leading order in $\epsilon$ or $1/N$, this integral is saturated at short distances,
\beq \int_0^{\infty} du \, u f_n(u) \to \frac{d_n}{\nu^{-1} - (D-2)} = \frac{d_n}{\nu_1}\eeq
and 
\beq \log{\frac{Z_n}{Z_1^n}} \approx - \frac{\pi n}{\nu_1} d_n (m L)^{d-1} \label{finalZn}\eeq
where we've dropped the constant contribution at the critical point $t = 0$. So, the universal coefficient $r_n$ of the correlation length correction, eq. (\ref{sndef}), is given by,
\beq r_n \approx - \frac{\pi n}{(1-n) \nu_1} d_n\label{renh}\eeq
Thus, to leading order the problem is reduced to evaluating the coefficient $d_n$ in (\ref{fshort}). Since this coefficient is a short distance property, we may work directly at the critical point. Note in particular that in the large $N$ limit, $d_n \sim O(N)$, so our result for $\log{\frac{Z_n}{Z_1^n}}$ scales as $N^2$. This is in contrast to the linear in $N$ behaviour that one would obtain at any finite order in the $1/N$ expansion for a fixed correlation length $\xi$.

It turns out that the leading term (\ref{renh}) behaves as $r_n \sim (n-1)$ for $n \to 1$ in both expansions and does not contribute to the entanglement entropy, eq. (\ref{SSn}). Thus, the correlation length correction to the entanglement entropy has the expected scaling $r \sim O(N)$. To proceed systematically beyond the leading order one needs to use renormalization group (RG) technology that will be developed explicitly in the context of $4-\epsilon$ expansion in section \ref{sec:RG}.

%

\section{$4-\epsilon$ expansion: correlation Length Correction}\label{sec:4eps}
In this section we compute the correlation length correction to the entanglement entropy in $4-\epsilon$ expansion. Recall that for the interacting $O(N)$ model, $\nu_1 = \nu^{-1} - (D-2) \sim O(\epsilon)$ in $D = 4-\epsilon$ dimensions, hence the argument in section \ref{sec:general} can be applied. This is also true for the non-interacting (Gaussian) fixed point for $D = 4-\epsilon$, where $\nu_1 = \epsilon$, allowing us to compare the predictions of our method to the exact calculations of Ref. \onlinecite{Cardy}. We first consider the Gaussian fixed point and then proceed to the Wilson-Fisher fixed point.
\subsection{Gaussian theory}
Consider the Gaussian theory,
\beq L = \frac{1}{2} (\d_{\mu} \phi)^2 + \frac{t}{2} \phi^2\label{Gaussian}\eeq
where,  $t = m^2$. We need to compute the expectation value,
\beq  \langle \phi^2(x) \rangle_n -   \langle \phi^2(x) \rangle_1\label{tphin}\eeq
at the critical point, $t = 0$. To leading order we may work in $D = 4$. The massless propagator on an $n$-sheeted Riemann-surface in $D = 4$ is known to be,\cite{Linet}
\beq G_n(r,r',\theta,x_\perp) = \frac{\sinh(\eta/n)}{8\pi^2 n r r' \sinh \eta (\cosh(\eta/n) - \cos(\theta/n))}\label{prop4D}\eeq
where
\beq \cosh \eta = \frac{r^2 +r'^2 + x^2_{\perp}}{2 r r'}\eeq
Hence,
\beq \langle \phi^2(x) \rangle_n -   \langle \phi^2(x) \rangle_1 = \frac{N}{48 \pi^2 r^2} \left(\frac{1}{n^2} - 1\right)\label{J0}\eeq
So comparing to eqs. (\ref{tphi2}), (\ref{fshort}), we obtain,
\beq d_n = \frac{N}{48 \pi^2} \left(\frac{1}{n^2} - 1\right),\quad\quad  \mbox{Gaussian fixed point}, \quad D = 4-\epsilon \label{Gaussiand}\eeq

We can now use eq. (\ref{renh}) to compute the coefficient $r_n$ of the correlation length correction. As noted above for the Gaussian theory, $\nu_1 = \epsilon$, so
\beq r_n = - \frac{N}{48 \pi \epsilon} \left(1+ \frac{1}{n}\right)\label{rnfree}\eeq
and for the entanglement entropy proper,
\beq r = \lim_{n \to 1} r_n = - \frac{N}{24 \pi \epsilon}\label{rfree}\eeq
This can be compared to the exact result of Ref. \onlinecite{Cardy},
\beq r_n = N \frac{\Gamma(\frac{2-D}{2})}{24 (4 \pi)^{(D-2)/2}} \left(1 + \frac{1}{n}\right)\label{ExactFreern}\eeq
Eq. (\ref{ExactFreern}) is in agreement with our result (\ref{rnfree}) to leading order in $\epsilon$, which is all that the discussion in section \ref{sec:general} guarantees.

\subsection{Interacting theory}\label{sec:int4eps}
We now proceed to consider the interacting $O(N)$ model, eq. (\ref{soft}).
We again need to compute the expectation value (\ref{tphin}). Naively, one would expect that at leading order in $\epsilon$, one can work with the mean-field approximation, $u = 0$, recovering the result (\ref{Gaussiand}). Then, one would simply substitute (\ref{Gaussiand}) into eq. (\ref{renh}) and use the appropriate $\nu_1$, eq. (\ref{nu14eps}), for the Wilson-Fisher fixed point. However, such reasoning turns out to be too simple minded, as it neglects ``boundary perturbations." Indeed, our conical singularity will generally induce local perturbations at $r = 0$. Of these, the term with the lowest engineering dimension is,
\beq \delta S = \frac{c}{2} \int d^{D-2} x_{\perp}\, \phi^2(r =0 ,x_\perp) \label{boundary}\eeq
In the absence of the conical singularity this perturbation is known to be irrelevant in the $O(N)$ model as the scaling dimension $[c] = \nu^{-1} - 2 < 0$.\cite{Vojta2000} However, as we will now show, the presence of the conical singularity will modify the renormalization group flow of the coefficient $c$.

\begin{figure}[h]
\begin{center}
\includegraphics[width=2.5in]{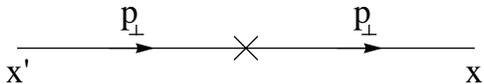}
\caption{Leading correction to the propagator $\delta {\cal G}^{1,0}$ due to the boundary perturbation. Here and below, a cross denotes an interaction vertex of $c$.}
\label{FigG10}
\end{center}
\end{figure}
\begin{figure}[h]
\begin{center}
\includegraphics[width=5in]{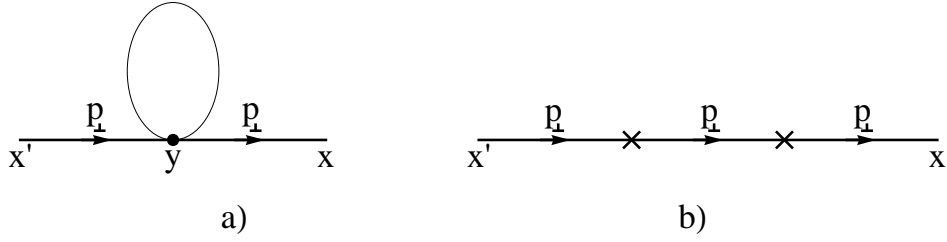}
\caption{Corrections to the propagator, a) $\delta {\cal G}^{0,1}$ and b) $\delta {\cal G}^{2,0}$. Here and below, a dot denotes an interaction vertex of $u$.}
\label{FigG0120}
\end{center}
\end{figure}
\begin{figure}[h]
\begin{center}
\includegraphics[width=7in]{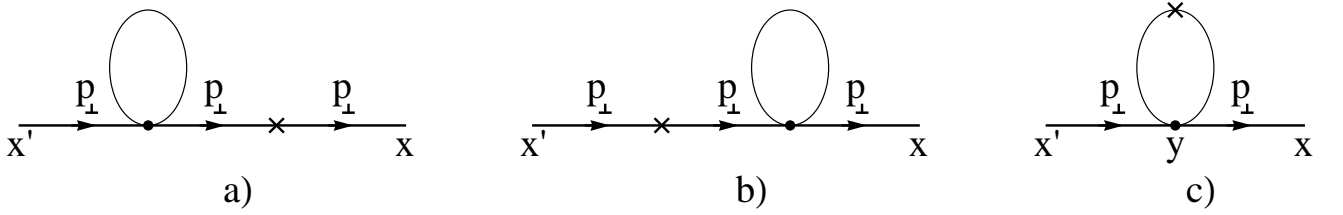}
\caption{Corrections to the propagator $\delta {\cal G}^{1,1}$.}
\label{FigG11}
\end{center}
\end{figure}
The engineering dimension of the coupling constant $c$ is zero in any space-time dimension $D$. 
We wish to compute the $\beta$-function, $\beta(c)$. Let us perform perturbation theory in $u$ and $c$ for the two-point function $\langle \phi_{\alpha}(x) \phi_{\beta}(x')\rangle = \delta_{\alpha \beta} {\cal G}(x,x')$. It is sufficient to work in $D = 4$ dimensions to compute the leading terms in $\beta(c)$. We use a mixed momentum/position $p_{\perp}$, $x_{\prl}$ representation.  To first order in $c$ and zeroth order in $u$, we have the simple diagram in Fig. \ref{FigG10},
\beq \delta^{1,0} {\cal G}(x_{\prl},x'_{\prl}, p_{\perp}) = - c \,G_n(x_{\prl},0,p_{\perp}) G_n(0, x'_{\prl}, p_{\perp}) \label{delta10}\eeq
where the superscripts on $\delta$ indicate the order in $c$ and $u$. Notice that the bare propagator $G_n(x,x')$, eq. (\ref{prop4D}), remains finite as its arguments approach the conical singularity. In fact,
\beq G_n(0,x) = \frac{1}{n} G_1(x)\label{GnG1}\eeq 
Also, $G_n(x_{\prl}, x'_{\prl}, p_{\perp})$ is just the two dimensional massive propagator $(-\nabla^2_2 + p^2_{\perp})^{-1}$ on an $n$-sheeted Riemann surface. In particular, $G_n(x_{\prl}, 0, p_{\perp}) = \frac{1}{n} K_0(p_{\perp} |x_{\prl}|)$ (which implies that the relation (\ref{GnG1}) is actually correct in any dimension).  Thus, the correction (\ref{delta10}) is finite. 

We next consider the Hartree-Fock (first order in $u$) correction to the propagator, Fig. \ref{FigG0120} a),
\beq \delta^{0, 1} {\cal G}(x_{\prl},x'_{\prl}, p_{\perp}) = - (N+2) u\int d^2 y_{\prl} \,G_n(x_{\prl}, y_{\prl}, p_{\perp}) G_n(y_{\prl}, x'_{\prl}, p_{\perp}) (G_n(y,y) - G_1(y,y)) \label{delta01}\eeq
We have already evaluated $G_n(y,y) - G_1(y,y) \sim \frac{1}{y^2_{\prl}}$, eq. (\ref{J0}). Thus, the integral (\ref{delta01}) has an ultraviolet divergence in the region $y_{\prl} \to 0$,
\beq \delta^{0, 1} {\cal G}(x_{\prl},x'_{\prl}, p_{\perp}) \stackrel{UV}{=} \frac{(N+2) u}{24 \pi} \left(n - \frac{1}{n}\right) G_n(x_{\prl},0,p_{\perp}) G_n(0, x'_{\prl}, p_{\perp}) \log(\Lambda)\eeq
Notice that this divergence is local to the conical singularity and, as is evident from eq. (\ref{delta10}), can be canceled by an additive renormalization of the coupling constant $c$. Hence, the perturbation (\ref{boundary}) will be automatically induced by the presence of the conical singularity.

We also consider the second order contribution in $c$ to the propagator, Fig. \ref{FigG0120} b),
\beq \delta^{2, 0} {\cal G}(x_{\prl},x'_{\prl}, p_{\perp})= c^2 G_n(x_{\prl}, 0, p_{\perp}) G_n(0, x'_{\prl}, p_{\perp}) G_n(0,0, p_{\perp})\eeq
The quantity $G_n(0, 0, p_{\perp})$ is UV singular,
\beq G_n(0,0, p_{\perp}) = \int d^2 y_{\perp} \,G_n(0,0,y_{\perp}) e^{-i p_{\perp} y_{\perp}} = \frac{1}{4 \pi^2 n} \int d^2 y_{\perp} \frac{1}{y^2_{\perp}} e^{i p_{\perp} y_{\perp}} \stackrel{UV}{=} \frac{1}{2 \pi n} \log(\Lambda/p_{\perp})\eeq
so
\beq \delta^{2, 0} {\cal G}(x_{\prl},x'_{\prl}, p_{\perp})\stackrel{UV}{=} \frac{c^2}{2 \pi n} G_n(x_{\prl}, 0, p_{\perp}) G_n(0, x'_{\prl}, p_{\perp}) \log(\Lambda) \label{delta20} \eeq
The divergence of (\ref{delta20}) is a manifestation of the well-known fact that the two-dimensional $\delta$-function potential requires regularization. Again, from (\ref{delta10}), we observe that the divergence can be eliminated by a renormalization of the coefficient $c$.

Finally, we consider corrections which are bilinear in $c$ and $u$, Fig. \ref{FigG11}. For $c$ - small, these corrections are generally subleading compared to $\delta^{0,1} {\cal G}$, Fig. \ref{FigG0120} a). However, for $n \to 1$, $\delta^{0, 1} {\cal G}$ vanishes, and the diagram in Fig. \ref{FigG11} c) becomes important. On the other hand, the diagrams in Figs. \ref{FigG11} a,b) can be ignored to leading order for all $n$ since they also vanish at $n = 1$.\footnote{Technically, these diagrams contain $(\log \Lambda)^2$ divergences, and one needs to use a consistent regularization method to evaluate them.} With this in mind, we only need to evaluate Fig. \ref{FigG11} c) at $n = 1$. We recognize, that this is just the diagram corresponding to the usual multiplicative renormalization of the $\phi^2$ operator. Explicitly,
\bea \delta^{1,1}{\cal G}(x_\prl,x'_\prl,p_{\perp}) &\stackrel{n=1}{=}& (N+2) u c\int d^2 y_{\prl}\, G_1(x_{\prl}, y_{\prl}, p_{\perp}) G_1(y_{\prl},x'_{\prl}, p_{\perp}) \int d^2 z_{\perp} \,G_1(y_{\prl},z_{\perp})^2 \nn\\ &=& (N+2) u c\int d^2 y_{\prl} \,G_1(x_{\prl}, y_{\prl}, p_{\perp}) G_1(y_{\prl},x'_{\prl}, p_{\perp})\frac{1}{16 \pi^3 y^2_{\prl}}\nn \\ &\stackrel{UV}{=}& \frac{(N+2) u c}{8 \pi^2} G_1(x_{\prl}, 0, p_{\perp}) G_1(0,x'_{\prl}, p_{\perp})\log \Lambda \eea

We can now introduce counterterms to cancel the divergences considered above,
\beq c = c_r + \left(\frac{(N+2) u_r}{24 \pi} \left(n -\frac{1}{n}\right) + \frac{(N+2)  u_r c_r}{8 \pi^2} + \frac{c^2_r}{2 \pi n}\right) \log(\Lambda/\mu)\eeq
where $c_r$ and $u_r$ are the renormalized coupling constants and $\mu$ is the renormalization scale. Note that the coefficient of the $u_r c_r$ term has been only computed at $n = 1$. So,
\beq \beta(c_r) = \mu \frac{\d}{\d \mu} c_r\Big|_{c,u} = \frac{(N+2) u_r}{24 \pi} \left(n-\frac{1}{n}\right)  + \frac{(N+2) u_r c_r}{8 \pi^2} + \frac{c^2_r}{2 \pi n} \label{betacr}\eeq
Note that the RG flow of $u$ is not affected by the boundary perturbation or by the presence of the conical singularity,
\beq \beta(u_r) = -\epsilon u_r + \frac{N+8}{8 \pi^2} u^2_r\eeq
and we have the usual Wilson-Fisher fixed point $u^* = \frac{8 \pi^2 \epsilon}{N+8}$.

\begin{figure}[h]
\begin{center}
\includegraphics[width=5in]{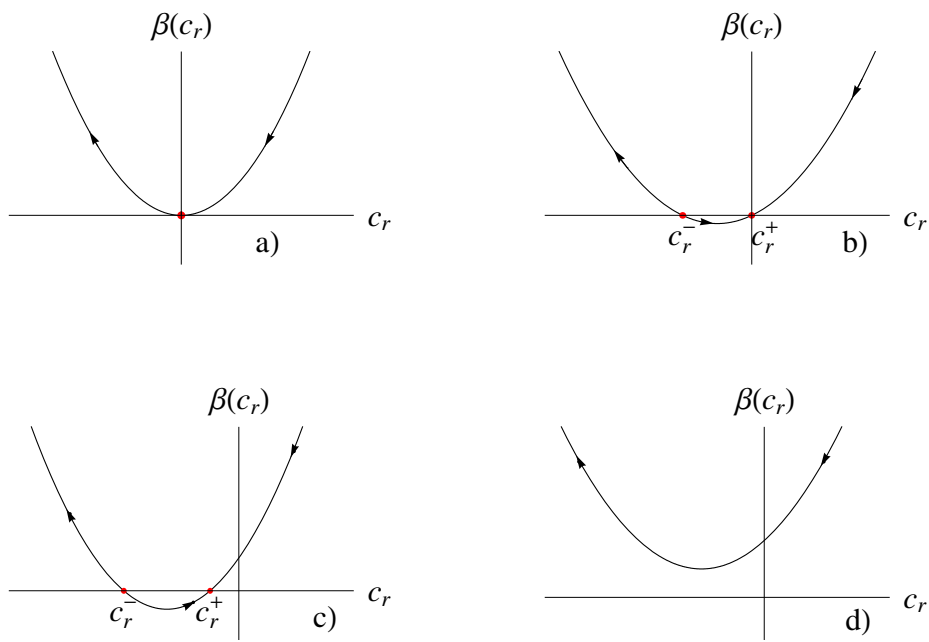}
\caption{$\beta$-function of the boundary coupling $c_r$ for a) Non-interacting theory ($u$ = 0), b) Interacting theory, $n = 1$, c) Interacting theory, $n < n_c$, d) Interacting theory, $n > n_c$.}
\label{Figbetac}
\end{center}
\end{figure}

We now discuss the RG flow of $c_r$ in detail. Let us start with the non-interacting theory, $u = 0$, which corresponds to the well-studied problem of a particle in a two-dimensional $\delta$-function potential. Then, $\beta(c_r) = \frac{1}{2 \pi n} c_r^2$. As demonstrated in Fig. \ref{Figbetac} a), the coupling constant $c_r$ flows logarithmically to zero for $c_r > 0$ and runs away to $- \infty$ for $c_r < 0$, signaling the formation of a bound state. 

Next, consider turning on the interaction $u$, in the absence of conical singularity ($n = 1$). Then, $\beta(c_r) = -\eta_2(u_r) c_r + \frac{c^2_r}{2 \pi}$, where $\eta_2$ is just the usual anomalous dimension of the $\phi^2$ operator, ($[\phi^2] = D-2 - \eta_2$),
\beq \eta_2(u_r) = - \frac{(N+2) u_r}{8 \pi^2}\eeq
The RG flow of $c$ is sketched in Fig. \ref{Figbetac} b). We find two fixed-points: $c^+_r = 0$ and $c^-_r =- \frac{N+2}{N+8} (2 \pi \epsilon)$. The first fixed point $c^+_r = 0$ is stable, due to $\beta'(c_r = 0) = -\eta_2(u^*) > 0$, which implies that for $c$ - small, the perturbation (\ref{boundary}) is irrelevant.\cite{Vojta2000} This conclusion can be immediately reached by consideration of scaling dimensions at the interacting fixed point, since $[c] = D-2 - [\phi^2] = \eta_2 < 0$.

The second fixed point $c^-_r$ is unstable, and for $c_r < c^-_r$ the RG flow runs away to $c_r = -\infty$. Naively, such a flow may be interpreted as a tendency of $\phi$ to condense in the vicinity of $r = 0$. However, this would result in a condensate that is effectively $D- 2 < 2$ dimensional, which, at least for $N \ge 2$ and $t > 0$, is prohibited by the Mermin-Wagner theorem. Exactly at the critical point, long-range forces could, in principle, stabilize the condensate. However, as we will discuss in section \ref{sec:largeN}, large-$N$ expansion suggests that no such condensation occurs even at $t=0$, and the flow actually terminates at a scale invariant fixed-point, which is inaccessible in our perturbative expansion. However, this fixed point can likely be interpreted in terms of a fluctuating ``boundary" order parameter.

Finally, we proceed to the interacting case in the presence of a conical singularity. For $n < n_c \approx 1+ \frac{3}{4} \frac{N+2}{N+8} \epsilon$ we again obtain two fixed points, Fig. \ref{Figbetac} c),
\beq c^{\pm}_r = \pi \left(-\frac{N+2}{N+8} n \epsilon \pm \sqrt{\left(\frac{N+2}{N+8}\right)^2 n^2 \epsilon^2 - \frac{2}{3} \frac{N+2}{N+8} (n^2-1)\epsilon}\right) \label{crpm}\eeq
The fixed point $c^+_r$ is stable, while $c^-_r$ is unstable. In the limit $n \to 1$, which is relevant for the computation of entanglement entropy, $c^+_r$ smoothly evolves to the $c^+_r = 0$ stable fixed point, which we obtained in the absence of the conical singularity. Moreover, for $n \to 1$, we expect the starting point of the RG flow $c_r \to 0$. Hence, for $n$ close to $1$ the RG flow will terminate at the fixed point $c^+_r$. Thus, the main effect of the conical singularity is to shift  $c^+_r$ away from $0$. The parametric magnitude of this shift depends on whether $1 - n \gg \epsilon$ or $|1-n| \ll \epsilon$:
\bea c^+_r &\approx& \pi \sqrt{\frac{2}{3} \frac{N+2}{N+8} (1-n^2) \epsilon}, \quad 1 -n \gg \epsilon \label{cfar1}\\
c^+_r &\approx& -\frac{2 \pi}{3} (n-1) - \frac{2 \pi}{9} \frac{N+8}{N+2} \frac{(n-1)^2}{\epsilon}, \quad |1-n|\ll \epsilon \label{cclose1}\eea
Thus, for $1-n \gg \epsilon$, $c^+_r \sim O(\sqrt{\epsilon})$: this is the regime in which the $u_r c_r$ term in the $\beta$-function (\ref{betacr}) can be ignored. On the other hand, for $|n-1| \ll \epsilon$, $c^+_r \sim (n-1) \ll \epsilon$ and the $u_r c_r$ term in $\beta(c_r)$ becomes important. Note that in both regimes, $c^+_r$ is parametrically small and the perturbative expansion in $c_r$ is justified.

For $n > n_c$, both fixed points disappear and the RG flow runs away to $c_r = -\infty$, Fig. \ref{Figbetac} d). As discussed above for the case $n = 1$, large $N$ analysis suggest that the flow is towards another fixed point (which itself evolves as a function of $n$). Now there are two possibilities. If as $n$ increases from $1$ to $n_c$, the initial value of $c_r$, determined by the microscopic details of the theory, satisfies $c_r(n) > c^-_r(n)$ then the run-off to the $c_r = -\infty$ fixed point will occur precisely at $n = n^* = n_c$. On the other hand, if the initial value of the coupling $c_r(n) < c^-_r(n)$ for $n > n^*$ where $1 < n^* < n_c$, the runaway to $c_r = -\infty$ will occur before $n$ reaches $n_c$. Note that the value of $n^*$ is generally non-universal. In either case, the long-distance physics is controlled by the $c^+_r$ fixed point for $n < n^*$ and the $c_r = -\infty$ fixed point for $n > n^*$. Thus, the constants $\gamma_n$, $r_n$, eqs. (\ref{gammandef}), (\ref{sndef}) will always have a discontinuity at some $n = n^*$, $1 < n^* \leq n_c$. 
Note that eqs. (\ref{gammandef}), (\ref{sndef}) are understood in the limit when the size of the regions whose entanglement entropy we are computing and the correlation length $\xi$ tend to infinity. However, as $n \to n^*$ a new divergent length scale emerges in the problem. In fact, we can think of the point $n = n^*$, $t = 0$ as a multicritical point. Thus, the limits $L$, $\xi \to \infty$ and $n \to n^*$ do not commute. In particular, if we fix $L$ or $\xi$, the dependence of the Renyi entropy on $n$ will be completely analytic. Moreover, the emergence of a new length-scale as $n \to n^*$ implies that the Renyi entropy in the cross-over region is not entirely universal. 



\begin{figure}[h]
\begin{center}
\includegraphics[width=4in]{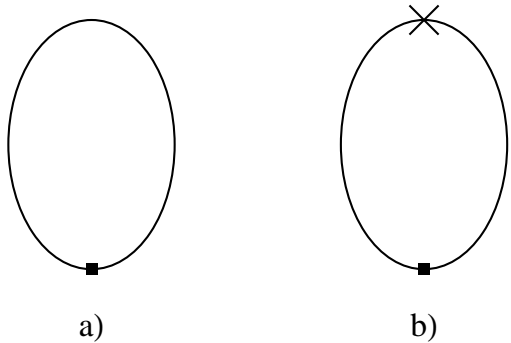}
\caption{Leading contributions to $\langle \phi^2(x)\rangle_n$ (denoted by a black square here and below): a) Mean-field result, b) Correction due to the boundary perturbation.}
\label{Figphi2c}
\end{center}
\end{figure}

Having discussed the non-trivial $n$-dependence of the Renyi entropy that occurs for $n$ away from $1$, we come back to the range $n < n_c$ and concentrate on the $c^+_r$ fixed point. We will from here on denote $c^+_r$ as $c^*_r$. Let us now compute the value of $\langle \phi^2(x)\rangle$ at this fixed point. The leading correction to the mean-field result, Fig. \ref{Figphi2c} a), eq. (\ref{J0}), is given by the diagram in Fig. \ref{Figphi2c} b),
\beq \delta^{1,0} \langle \phi^2(x) \rangle = - N c_r \int d^{D-2} y_{\perp} G^2_n(x,y) = - \frac{N c_r}{16 \pi^3 n^2} \frac{1}{r^2}\eeq
Since to leading order we still have $t = m^2$, from eqs. (\ref{tphi2}) and (\ref{fshort}),
\beq d_n \approx N \left[\frac{1}{48 \pi^2} \left(\frac{1}{n^2}-1\right) - \frac{c^+_r}{16 \pi^3 n^2}\right]\label{dn4eps}\eeq
and from eqs. (\ref{nu14eps}), (\ref{renh}), the coefficient of the correlation length correction to the Renyi entropy is,
\beq r_n  \approx - \frac{\pi n (N+8)}{6 \epsilon (1-n)} d_n\eeq 
As we see, in the regime $1 - n \gg \epsilon$, taking the boundary perturbation into account only weakly modifies the mean-field result for $d_n$, eq. (\ref{Gaussiand}), by a term of order $\sqrt{\epsilon}$. Note that $r_n$ is still strongly modified due to a different value of $\nu_1$. 

However, in the regime $|1-n| \ll \epsilon$, 
\beq d_n \approx \frac{N (N+8)}{ (N+2)}\frac{(n-1)^2}{72 \pi^2 \epsilon}, \quad |1-n| \ll \epsilon \label{dclose1}\eeq
\beq r_n \approx \frac{N (N+8)^2}{N+2} \frac{n-1}{432 \pi \epsilon^2},  \quad |1-n| \ll \epsilon \label{rclose1}\eeq
Thus, for $n \to 1$, the behavior of $d_n$ at the Wilson-Fisher is drastically different from the mean-field result, eq. (\ref{Gaussiand}). In particular, notice that to the present order in $\epsilon$, the correction due to the boundary perturbation precisely cancels the term linear in $n-1$ coming from eq. (\ref{J0}). The technical reason for this remarkable cancellation is as follows. For $n \to 1$, we expect $c_r \sim O(n-1)$, and we can work just to first order in $c$. Then, in considering the corrections to the propagator, we can drop the diagram in Fig. \ref{FigG0120} b), keeping only Figs. \ref{FigG0120} a) and \ref{FigG11} c). These diagrams are, essentially, Hartree-Fock corrections to the propagator, and the ``Hartree-Fock potential" at $y$ is just $\langle \phi^2(y) \rangle_n - \langle \phi^2(y)\rangle_1 \sim 1/y^2_{\prl}$. As a result, the diagrams diverge for $y_\prl \to 0$. The $\beta$-function for the coupling constant $c_r$ vanishes precisely when this divergence is absent, i.e. $\langle \phi^2(y) \rangle_n - \langle \phi^2(y)\rangle_1 = 0$.

The crucial consequence of eq. (\ref{dclose1}) is that to this order the correction to entanglement entropy proper, $r = \lim_{n \to 1} r_n = 0$. Thus,
\beq r \sim O(1), \quad D =4-\epsilon\eeq
We conclude that the correlation length dependent contribution to the entanglement entropy at the Wilson-Fisher fixed point is parametrically smaller than at the Gaussian fixed point in $D = 4-\epsilon$, eq. (\ref{rfree}). As a result, we have to proceed to higher order in $\epsilon$ to evaluate it. This will be done in the next section.

Before we perform the higher order computation, let us ask how do the correlation functions of the field $\phi(x)$ behave as $x$ approaches the conical singularity. This question is connected to the effective boundary conditions on the field $\phi$ that are generated at the singularity. In accordance with the general theory of boundary critical phenomena,\cite{Diehl} we expect the field $\phi$ to satisfy the operator product expansion (OPE),
\beq \phi(x_{\prl}, x_{\perp}) \sim r^{\alpha} \phi(0, x_{\perp}), \quad r \to 0\eeq
where $\phi(0,x_{\perp})$ is an operator living on the conical singularity. The exponent $\alpha$ can be extracted from the two-point function ${\cal G}(x,x')$. Combining the free propagator with the boundary correction, eq. (\ref{delta10}),
\beq {\cal G}(x_{\prl}, x'_{\prl}, p_{\perp}) \stackrel{x_\prl \to 0}{=} \left(1 + \frac{c_r}{2 \pi n} \log(p_{\perp} r)\right) G_n(0,x'_{\prl}, p_{\perp})\eeq
from which we conclude,
\beq \alpha = \frac{c^*_r}{2 \pi n}\label{alpha4eps}\eeq
Note that from eq. (\ref{crpm}) the exponent $\alpha$ is positive for $n < 1$, implying effective Dirichlet boundary conditions on $\phi(x)$ at the conical singularity. On the other hand, $\alpha$ is negative for $1 < n  < n_c$ and correlation functions of $\phi(x)$ exhibit a power-law divergence as $x_{\prl}$ approaches the origin.

\subsection{Beyond the leading order in $\epsilon$}

\subsubsection{The inhomogeneous renormalization group equation}\label{sec:RG}
At leading order in $\epsilon$, our calculation has relied on the integral in eq. (\ref{ffinal}) being saturated at short distances, $u = m r \to 0$, allowing us to work directly at the critical point. However, we saw that the coefficient $d_n$ of the short-distance asymptotic of $f_n$, eq. (\ref{fshort}), behaved as $d_n \sim (n-1)^2/\epsilon$ for $n \to 1$, giving no contribution to the entanglement entropy. We expect that to next order in $\epsilon$, $d_n$ will acquire a term linear in $n-1$, $d_n \sim \epsilon (n-1)$, which by eq. (\ref{finalZn}) will give a contribution of $O(1)$ to $S$. Notice that this is of the same order as the contribution of the long distance, $u \to \infty$, part of the integral (\ref{ffinal}), which now has to be taken into account. Thus, we need to compute the long distance part of $f_n$ to leading order in $\epsilon$ and the short distance part to subleading order. Although the separation between short and long distance contributions is unambiguous to present order, it is convenient to introduce a formalism that allows one to consistently treat the problem order by order in $\epsilon$.\footnote{We note that the discussion below closely parallels the renormalization group technology used to calculate the specific heat in the classical $O(N)$ model.} 

Let us define,
\beq \Phi(p) = n \int_{1-\mathrm{sheet}} d^2 x_{\prl} \, \left(\big\langle \left[\phi^2(x)\right]_r\big\rangle_n - \big\langle \left[\phi^2(x)\right]_r\big\rangle_1\right) e^{-i \vec{p} \vec{x}}\label{Phi}\eeq
Here, we have introduced the usual renormalization of the $\phi^2$ operator,
\beq [\phi^2(x)]_r = \frac{Z_2}{Z} \phi^2(x), \quad t_r = \left(\frac{Z_2}{Z}\right)^{-1} t\eeq
We are considering $\Phi$ at a finite momentum $p$ in order to make $\Phi$ well-defined even at the critical point, $t = 0$. We are actually interested in computing $\Phi$ at $p = 0$ in the gapped phase, $t \neq 0$, as from eq. (\ref{ddt}),
\beq t_r \frac{\d}{\d t_r} \log \frac{Z_n}{Z^n} = - \frac{1}{2} t_r \Phi(p =0) L^{D-2} \label{trdtr}\eeq

As already observed in section \ref{sec:general}, although the integrand in (\ref{Phi}) is finite, the integral diverges logarithmically for $|x| \to 0$ at each order in $u$. Thus, $\Phi(p)$ requires an additive renormalization,
\beq \Phi(p) = \Phi_r(p) + C(u_r,c_r,\mu/\Lambda) \mu^{-\epsilon} \label{addren}\eeq
where $C$ is a renormalization constant. We will use dimensional regularization  below, so that $C$ is, in fact, just a function of $u_r$ and $c_r$. Then $\Phi_r$ satisfies the inhomogeneous renormalization group equation,
\beq \left(\mu\frac{\d}{\d \mu} + \beta(u_r) \frac{\d}{\d u_r} + \beta(c_r) \frac{\d}{\d c_r} - \eta_2(u_r) \left(1 + t_r \frac{\d}{\d t_r}\right)\right) \Phi_r = B(u_r, c_r) \mu^{-\epsilon} \label{RGinhom}\eeq
with
\beq B(u_r, c_r) = - \left(\beta(u_r) \frac{\d}{\d u_r} + \beta(c_r) \frac{\d}{\d c_r} - (\eta_2(u_r) + \epsilon)\right) C(u_r,c_r)\label{B}\eeq
where as usual,
\beq \eta_2(u_r) = \mu\frac{\d}{\d \mu}\bigg|_{u} \log \frac{Z_2}{Z}\eeq
Note that $B$ must be finite, as the left hand side of eq. (\ref{RGinhom}) is finite. The solution to (\ref{RGinhom}) can be represented as a sum of the solution to the homogeneous RG equation and a particular solution. In the scaling limit, $t_r \to 0$,
\beq \Phi_r(p = 0) = A_s \mu^{-\epsilon} \left(\frac{t_r}{\mu^2}\right)^{-(\epsilon+\eta_2)/(2 + \eta_2)} + A_{ns}(u_r,c_r)\mu^{-\epsilon} \label{RGsol}\eeq 
where the coefficient of the particular solution $A_{ns}$ satisfies,
\beq \left(\beta(u_r) \frac{\d}{\d u_r} + \beta(c_r) \frac{\d}{\d c_r} - (\eta_2(u_r) + \epsilon)\right) A_{ns}(u_r, c_r) = B(u_r,c_r)\eeq
Hence, at the critical point, 
\beq A_{ns}(u^*_r, c^*_r) = -\frac{1}{\eta_2 + \epsilon} B_* = -\frac{1}{\nu_1} B_* \label{Ans}\eeq
where we recall our definition in section \ref{sec:general}, $\nu_1 = \nu^{-1}-(D-2)$ and $\nu^{-1} = 2 + \eta_2$. 

Thus, from eq. (\ref{trdtr}),
\beq \log\frac{Z_n}{Z^n} = - \frac{A_s}{2 \nu (D-2)} \left(\mu \left(\frac{t_r}{\mu^2}\right)^{\nu}\right)^{D-2} L^{D-2} \eeq
where we've dropped terms analytic in $t_r$. Note that the mass gap $m$ is related to $\mu \left(\frac{t_r}{\mu^2}\right)^{\nu}$ via a finite proportionality constant, which at leading order in $\epsilon$ is just $1$. So to leading order,
\beq r_n \approx - \frac{A_s}{2 (1-n)} \label{ZAs}\eeq
Hence, we must compute $A_s$. To do so, we perturbatively calculate $\Phi_r(p=0)$ and $B(u_r,c_r)$. $A_s$ can then be determined by matching the perturbative expansion with the solution to the RG equation (\ref{RGsol}) a the critical point, where the corrections to scaling vanish. Notice that we always need to compute $B$ to one higher order in $\epsilon$ than $\Phi_r(p=0)$ due to the factor $\nu_1$ in the denominator of eq. (\ref{Ans}). Moreover, since $\Phi_r$ is finite for $\epsilon \to 0$, while $A_{ns} = - B_*/\nu_1$ behaves as $1/\epsilon$, to leading order $A_s = - A_{ns} = B_*/\nu_1$. Precisely this fact was utilized in section \ref{sec:general}, and we identify to leading order $B_* = 2 \pi n d_n$. 

\subsubsection{Regularization}\label{sec:reg}
For the purpose of computing the entanglement entropy $S$ we can work to linear order in $n-1$. Since the fixed point value $c_* \sim O(n-1)$, we also work to linear order in $c$. Therefore, all diagrams that include an insertion of $c$ can be evaluated at $n = 1$. In addition, power counting indicates that if we work to linear order in $c$, all diagrams will be finite for $D < 4$ (by contrast, higher order diagrams in $c$, such as Fig. \ref{FigG0120} b) diverge even for $D < 4$). Thus, we use dimensional regularization and minimal subtraction below. We remind the reader that in dimensional regularization the bare coupling constant $u = \mu^{\epsilon} u_r Z_u/Z^2$. We list below the renormalization constants in the MS scheme to the order that they will be needed in our calculation.
\bea \frac{Z_u}{Z^2} &=& 1 + \frac{(N+8)}{\epsilon} \frac{u_r}{8 \pi^2}\\
\frac{Z_2}{Z} &=& 1 +  \frac{(N+2)}{\epsilon} \frac{u_r}{8 \pi^2} + \frac{(N+2)(N+5)}{\epsilon^2}\left(\frac{u_r}{8\pi^2}\right)^2 - \frac{5(N+2)}{4 \epsilon}\left(\frac{u_r}{8\pi^2}\right)^2 \label{Z2}\eea
Correspondingly,
\bea \beta(u_r) &=& - \epsilon u_r + \frac{(N+8) u^2_r}{8 \pi^2}\\
\eta_2(u_r) &=& - (N+2) \frac{u_r}{8 \pi^2} \left(1 - \frac{5}{2} \frac{u_r}{8 \pi^2}\right)\eea

As we saw, the boundary coupling constant $c$ will also require renormalization. To linear order in $c$,
\beq c = D(u_r) + \frac{Z_2}{Z} c_r \label{crgen}\eeq
where we observe that the multiplicative renormalization of $c$ to zeroth order in $(n-1)$ is just $Z_2/Z$.  On the other hand, the additive renormalization, which behaves as $D(u_r) \sim (n-1)$ for $n \to 1$, needs to be computed explicitly. So the $\beta$-function,
\beq \beta(c_r) = - \left(\frac{Z_2}{Z}\right)^{-1} \beta(u_r) \frac{\d D}{\d u_r} - \eta_2(u_r) c_r \label{betacgen}\eeq

\subsubsection{Entanglement entropy to $O(1)$}
To calculate the entanglement entropy to $O(1)$ in $\epsilon$, we need to find the finite part of $\Phi(p=0)$, eq. (\ref{Phi}), at $t \neq 0$ to $O(1)$ in $u$ and the divergent part of $\Phi(p)$, which determines $B$, eq. (\ref{B}), to $O(u)$. 

$\Phi(p)$ to $O(1)$ in $u$ is given by the two diagrams in Fig. \ref{Figphi2c}. The diagram Fig. \ref{Figphi2c} a) is just the mean field contribution computed in Ref. \onlinecite{Cardy},
\bea \Phi(p=0)_{MF} &=& N \int d^2 x_{\prl} \, \left(G_n(x,x) - G_1(x,x)\right) \nonumber \\
&=& N \int \frac{d^{D-2} k_{\perp}}{(2 \pi)^{D-2}} \int d^2 x_{\prl} \left(G^{D=2}_n(x,x;k^2_{\perp} + m^2) -n\to 1\right)\nn\\ &=& - \frac{N}{12} \left(n-\frac{1}{n}\right) \int \frac{d^{D-2} k_{\perp}}{(2 \pi)^{D-2}} \frac{1}{k^2_{\perp} + m^2} \nonumber \\
&=& - \frac{N}{12} \left(n-\frac{1}{n}\right) \frac{\Gamma(2-D/2)}{(4\pi)^{D/2-1}} m^{D-4} \label{Phim1}\eea
where $G^{D=2}_n(x,x';M^2)$ is the two dimensional massive propagator on the $n$-sheeted Riemann surface, and we have used the result proved in Ref. \onlinecite{Cardy},
\beq \int d^2 x_{\prl} \left(G^{D=2}_n(x,x;M^2) -G^{D=2}_1(x,x;M^2)\right) = - \frac{1}{12} \left( n - \frac{1}{n}\right) \frac{1}{M^2} \label{CardyD2int}\eeq

The diagram in Fig. \ref{Figphi2c} b) is the boundary correction,
\beq \delta^{1,0}\Phi(p=0) = - N c_r \int d^2 x_{\prl} \int d^{D-2} y_{\perp} G^2_1(x_{\prl}, y_{\perp}) = - N c_r \frac{\Gamma(2-D/2)}{(4 \pi)^{D/2}} m^{D-4}\label{Phim2}\eeq

Combining eqs. (\ref{Phim1}), (\ref{Phim2}),
\beq \Phi(p=0) \stackrel{O(1)}{=} - N \left(\frac{n-1}{12 \pi} + \frac{c_r}{8 \pi^2}\right) \left(\frac{1}{\epsilon} + \frac{1}{2} \log 4 \pi - \frac{\gamma}{2} - \log(m/\mu)\right) \mu^{-\epsilon}\eeq
where we keep only terms linear in $n-1$.

Subtracting the pole, we obtain for the additive renormalization constant $C$, eq. (\ref{addren}),
\beq C \stackrel{O(1)}{=} - N \left(\frac{n-1}{12 \pi} + \frac{c_r}{8 \pi^2}\right) \frac{1}{\epsilon} \label{CO1}\eeq
and consequently from eq. (\ref{B}),
\beq B \stackrel{O(1)}{=} \epsilon C = - N \left(\frac{n-1}{12 \pi} + \frac{c_r}{8 \pi^2}\right)\label{BO1}\eeq
and
\beq \Phi_r(p=0) \stackrel{O(1)}{=} - N \left(\frac{n-1}{12 \pi} + \frac{c_r}{8 \pi^2}\right) \left(\frac{1}{2} \log 4 \pi - \frac{\gamma}{2} - \log(m/\mu)\right) \mu^{-\epsilon}\eeq
In particular, at the critical point, by eq. (\ref{cclose1}),
\beq c^*_r \stackrel{O(1)}{=} - \frac{2 \pi}{3} (n-1)\eeq
and
\beq \Phi^*_r(p = 0) = O(\epsilon), \quad B_* = O(\epsilon)\eeq
Thus, in the minimal subtraction scheme $\Phi^*_r(p=0)$ vanishes at the critical point to $O(1)$ in $\epsilon$. The fact that $B_* = 2 \pi n d_n$ vanishes to $O(1)$ in $\epsilon$ has already been observed in section \ref{sec:int4eps}. Thus, from eqs. (\ref{RGsol}), (\ref{Ans}),
\beq A_s \stackrel{O(1)} = \frac{B_*}{\nu_1}\label{AsB}\eeq

We now proceed to evaluate $B$ to $O(\epsilon)$. To do this, we compute $\Phi(p)$ at the critical point. We first evaluate $\langle [\phi^2]_r\rangle_n - \langle [\phi^2]_r\rangle_1$ and use it to determine the renormalization of the coupling $c$ in dimensional regularization. We then perform the Fourier transform, eq. (\ref{Phi}), to find the subtraction constant $C$ and hence $B$. To leading order, we have the two familiar diagrams in Fig. \ref{Figphi2c},
\beq \langle \phi^2(x) \rangle_n - \langle \phi^2(x) \rangle_1 \stackrel{O(1)}{=} N \left[J(D) - c_r \frac{\Gamma(D/2-1)^3}{16 \pi^{D/2+1}\Gamma(D-2)}\right] \frac{1}{r^{D-2}}\eeq
where we've defined,
\beq G_n(x,x) - G_1(x,x) = \frac{J(D)}{r^{D-2}}\label{GJ}\eeq
Note that in dimensional regularization $\langle \phi^2 \rangle_1 = N G_1(x,x) = 0$ at the critical point. We will show in section \ref{sec:betacu2} that to linear order in $n-1$,
\beq J(D) = (n-1)  \frac{\Gamma(D/2)^3}{4 \pi^{D/2} (1-D/2) \Gamma(D)}\label{JD}\eeq
In particular, $J(D=4) = - \frac{n-1}{24 \pi^2}$ in agreement with eq. (\ref{J0}). We note that the diagrams that contain the tadpole (\ref{GJ}) can effectively be evaluated with $n = 1$. The computation is simplest in position space, where one uses,
\beq G_1(x,x') = \frac{\Gamma(D/2-1)}{4 \pi^{D/2} |x-x'|^{D-2}} \label{G1D}\eeq

\begin{figure}[h]
\begin{center}
\includegraphics[width=4in]{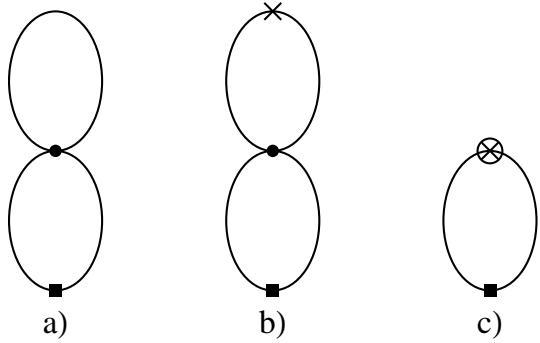}
\caption{Contributions to $\langle \phi^2(x)\rangle_n - \langle \phi^2(x)\rangle_1$ at order $u$. The counterterm $\delta^1 c$ is denoted by a circled cross here and below.}
\label{Figphi2u}
\end{center}
\end{figure}

At order $u$, $\langle [\phi^2]\rangle_n - \langle [\phi^2]\rangle_1$ receives contributions from the diagrams in Fig. \ref{Figphi2u}. Note that the diagram c) is the renormalization of the coupling constant $c_0 = c_r + \delta^1 c + ...$. Taking the multiplicative renormalization of the operator $\phi^2$ into account, we obtain,
\bea \langle [\phi^2]_r\rangle_n - \langle [\phi^2]_r\rangle_1 &\stackrel{O(u)}{=}& N \left(\frac{Z_2}{Z}\frac{1}{r^{D-2}} - \frac{\Gamma(D/2-1) \Gamma(2-D/2)^2}{16 \pi^{D/2} (D-3) \Gamma(4-D)} \frac{(N+2) u_r \mu^{\epsilon}}{r^{2(D-3)}}\right) \nonumber \\
&\times& \!\! \!\!\!\! \left(J-\frac{\Gamma(D/2-1)^3}{16 \pi^{D/2+1} \Gamma(D-2)} c_r\right) - N \frac{\Gamma(D/2-1)^3}{16 \pi^{D/2+1} \Gamma(D-2)} \frac{\delta^1 c}{r^{D-2}}\label{phi2u}\eea
Performing minimal subtraction,
\beq \delta^1 c = \frac{(N+2) u_r}{\epsilon} \left(\frac{n-1}{12 \pi} + \frac{c_r}{8 \pi^2}\right) \label{delta1c}\eeq
Notice that the coefficient of the multiplicative renormalization is precisely $Z_2/Z$ as expected. We also obtain the additive renormalization constant, eq. (\ref{crgen}), 
\beq D(u_r) =  \frac{(N+2) u_r}{\epsilon} \frac{n-1}{12 \pi}\eeq
Hence, from eq. (\ref{betacgen}), to first order in $u$,
\beq \beta(c_r) \stackrel{O(u)}{=} (N+2) u_r \left(\frac{n-1}{12 \pi} + \frac{c_r}{8 \pi^2}\right)\eeq
in agreement with the expression (\ref{betacr}) obtained earlier using cut-off regularization.

By Fourier transforming eq. (\ref{phi2u}), we can compute $\Phi(p)$ at the critical point to order $u$. From the divergent part, we obtain the additive renormalization constant $C$ (\ref{addren}),
\beq C(u_r,c_r) = -N \left(\frac{1}{\epsilon} + \frac{N+2}{\epsilon^2} \frac{u_r}{8 \pi^2}\right) \left(\frac{n-1}{12 \pi} + \frac{c_r}{8 \pi^2}\right) \label{COu}\eeq
which gives the $O(u)$ correction to our previous result (\ref{CO1}). Substituting  into eq. (\ref{B}), we obtain
\beq B \stackrel{O(u)}{=} - N \left(\frac{n-1}{12 \pi} + \frac{c_r}{8 \pi^2}\right)\eeq 
Comparing the above result to eq. (\ref{BO1}), we observe that $B$ receives no additional contributions at $O(u)$. Thus, from eq. (\ref{AsB}), 
\beq A_s \stackrel{O(1)}{=} - \frac{N (N+8)}{6 \epsilon} \left(\frac{n-1}{12 \pi} + \frac{c^*_r}{8 \pi^2}\right) \label{Asc}\eeq
which, upon determination of $c^*_r$ to order $\epsilon$ would yield the entanglement entropy, eq. (\ref{ZAs}).

\subsubsection{$\beta(c_r)$ to order $u^2$}\label{sec:betacu2}

\begin{figure}[h]
\begin{center}
\includegraphics[width=5in]{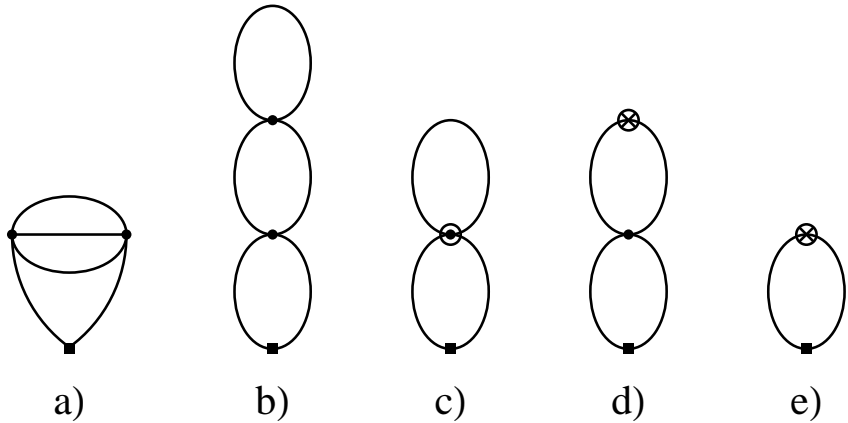}
\caption{Contributions to $\langle \phi^2(x)\rangle_n - \langle \phi^2(x)\rangle_1$ at order $u^2$ (diagrams involving insertions of $c_r$ are not shown). The counterterm for the coupling $u$ is shown as a circled dot.}
\label{Figphi2u2}
\end{center}
\end{figure}

To complete our calculation, we need the value of the fixed point coupling $c^*_r$ to order $\epsilon$. This requires the knowledge of $\beta(c_r)$ to order $u^2$. As before, we will determine the renormalization of $c$ by computing the expectation value $\langle [\phi^2]_r\rangle_n - \langle [\phi^2]_r\rangle_1$. 
As explained in section \ref{sec:reg}, we need to find only the additive renormalization of $c$. Hence, we ignore all diagrams with vertices proportional to $c_r$. At order $u^2$, we obtain the graphs shown in Fig. \ref{Figphi2u2}. 

Now we are faced with a new technical difficulty. Up to this point, to linear order in $n-1$, the conical singularity entered our calculations through the tadpole term $G_n(x,x) - G_1(x,x)$, whose form was fixed by dimensional analysis, eq. (\ref{GJ}), up to an overall constant $J(D)$. Moreover, the  renormalization constants only depended on $J(D=4)$, which could be extracted from the explicit form of the propagator, eq. (\ref{prop4D}). However, at the present order, we are faced with the diagram in Fig. \ref{Figphi2u2} a), which requires the full position dependence of the propagator $G_n(x,x')$. Yet, as far as we know, there is no simple expression for $G_n(x,x')$ in arbitrary dimension, and even in $D=4$ eq. (\ref{prop4D}) is rather awkward to work with. 

To address this problem, we expand the propagator $G_n(x,x')$ to linear order in $n-1$ in terms of the usual propagators $G_1(x,x')$, eq. (\ref{G1D}). The simplest way to do this is to consider the $O(N)$ model in the presence of an arbitrary metric $g_{\mu \nu}$,
\beq S = \int d^D x \sqrt{\det g} \left(g^{\mu \nu} \d_{\mu} \phi \d_{\nu} \phi + \frac{t}{2} \phi^2 + \frac{u}{4} \phi^4\right) \label{Sg}\eeq
It is convenient to parameterize the $n$-sheeted Riemann surface using rescaled variables,
\beq \tilde{r} = \sqrt{n} r, \quad \varphi = \theta/n\eeq
Then, the angular variable $\varphi \sim \varphi + 2 \pi$. We may also define,
\beq \tilde{\tau} = \tilde{r} \cos \varphi, \quad
\tilde{\mbox{x}} = \tilde{r} \sin \varphi\eeq
The coordinates $(\tilde{\tau}, \tilde{\mbox{x}})$ form the usual two dimensional Euclidean plane and uniquely specify each point on the
Riemann surface. With this choice of variables, the metric (\ref{metric}) in the $x_\prl$ plane becomes,
\beq g_{\alpha \beta} = n \delta_{\alpha \beta} + \left(\frac{1}{n} - n\right) \frac{\tilde{x}_\alpha \tilde{x}_\beta}{\tilde{x}^2}\eeq
where $\alpha$,$\beta$ run over $\tilde{\tau}$, $\tilde{\mbox{x}}$. Note that we have chosen to rescale $r$ in such a way that,
\beq \det g = 1\eeq
Moreover, expanding $g$ in powers of $n-1$, $g_{\alpha \beta} = \delta_{\alpha \beta} + \delta g_{\alpha \beta}$,
\beq \delta g_{\alpha \beta} \approx (n-1) \left(\delta_{\alpha \beta} -\frac{2 \tilde{x}_{\alpha} \tilde{x}_{\beta}}{\tilde{x}^2}\right)\eeq
We drop the tildes on variables $\tau, x$ in what follows. We can now obtain the usual Feynman graph expansion for the theory (\ref{Sg}), treating $\delta g_{\alpha \beta}$ as a perturbation. Note that all the integrals in the resulting expansion are over the usual $D$-dimensional Euclidean space. 
In particular, note that the bare propagator becomes,
\beq G_n(x,x') \approx G_1(x,x') + \delta G_n(x,x')\eeq
\beq \delta G_n(x,x') = (n-1) \int d^D y \, \left(\delta_{\alpha \beta} - \frac{2 y_{\alpha} y_{\beta}}{y_{\prl}^2}\right)\d_{\alpha}G_1(x-y) \d_{\beta} G_1(x'-y)  \label{Gnlin}\eeq
By performing the integral, we immediately obtain eq. (\ref{JD}) for $G_n(x,x) - G_1(x,x)$. 

Using the expansion (\ref{Gnlin}) we compute the divergent part of the diagrams in Fig. \ref{Figphi2u2} to linear order in $n-1$. After accounting for the multiplicative renormalization of the $\phi^2$ operator, eq. (\ref{Z2}), we extract the additive renormalization of the coupling constant $c$, eq. (\ref{crgen}) to $O(u^2)$,
\beq D(u_r) =  \frac{n-1}{12 \pi} \left( \frac{(N+2) u_r}{\epsilon} + \frac{(N+2) (N+5)}{\epsilon^2}\frac{u^2_r}{8 \pi^2} - \frac{7 (N+2)}{4 \epsilon} \frac{u^2_r}{8\pi^2}\right)\eeq
and from eq. (\ref{betacgen}),
\beq \beta(c_r) = (N+2) u_r \left(1- \frac{7}{2} \frac{u_r}{8 \pi^2}\right) \frac{n-1}{12 \pi} + (N+2) \frac{u_r}{8 \pi^2} \left(1 - \frac{5}{2} \frac{u_r}{8 \pi^2}\right) c_r\eeq
Hence,
\beq c^*_r = - \frac{2 \pi}{3}  \left(1- \frac{u^*_r}{8 \pi^2}\right)(n-1) = - \frac{2 \pi}{3} \left(1- \frac{\epsilon}{N+8}\right)(n-1) \label{cstar}\eeq
and from eq. (\ref{Asc}),
\beq A_s = - \frac{N}{72 \pi} (n-1)\eeq
which by eq. (\ref{ZAs}) finally yields the coefficient of the correlation length correction to the entanglement entropy,
\beq r = - \frac{N}{144 \pi}\eeq

\section{$4- \epsilon$ expansion: Finite size correction}\label{sec:4epsL}
In this section we compute the geometric corrections $\gamma$, $\gamma_n$  to the entanglement entropy and the Renyi entropy, eqs. (\ref{gammadef}), (\ref{gammandef}), at the critical point. 

As before, we consider two semi-infinite regions $A$ and $B$ with a boundary at $\mbox{x} = 0$. However, we now take the remaining $D-2$ spatial directions to have a finite length $L$. In order to avoid dealing with the zero mode, we use twisted boundary conditions along these directions. 
\beq \phi(x+L \hat{n}_i) = e^{i \varphi_i} \phi(x)\eeq
where $\hat{n}_i$ are unit vectors along the boundary. If the fields $\phi$ are real, then $\varphi_i = 0$ or $\pi$. On the other hand, in an $O(N)$ model with $N$ even, we can group our fields into $N/2$ complex pairs - then, an arbitrary twist is allowed (however, this breaks the $O(N)$ symmetry down to $U(1)\times SU(N/2)$). We note that when accessing $D = 3$ via $4-\epsilon$ expansion, we will choose all $\varphi_i$'s to be equal. 

Thus, the boundary between regions $A$ and $B$ is a $D-2$ dimensional torus. Since this manifold is smooth we expect the constants $\gamma$, $\gamma_n$ to be universal. Moreover, we don't have to take into account divergences which appear as $D \to 4$ when the boundary has a finite curvature,\cite{Ryu2} since this manifold is flat.

\subsection{Gaussian theory}
Let us begin with the free theory. We wish to compute,
\bea \log\frac{Z_n}{Z^n} &=& -\frac{N}{2} (Tr \log(-\d^2)_n  - n Tr \log(-\d^2)_1)\\
&=& -\frac{N}{2} \sum_{\vec{k}_{\perp}} \Big[Tr_{\parallel} \log(-\d^2_{\parallel} + \vec{k}^2_{\perp})_n  - n Tr_{\parallel} \log(-\d^2_{\parallel} + \vec{k}^2_{\perp})_1\Big]
\label{Tr}
\eea
where $k^i_{\perp} = \frac{2 \pi n_i + \varphi_i}{L}$ and $n_i$ are integers. We leave the regularization of eq. (\ref{Tr}) implicit for now (we will later use dimensional regularization). Eq. (\ref{Tr}) involves the partition function of the two-dimensional massive gaussian theory evaluated in  Ref. \onlinecite{Cardy},
\beq \log\frac{Z_n}{Z^n}\Big|_{D=2} = -\frac{1}{2} (Tr_{\parallel} \log(-\d^2_{\parallel} + m^2)-n Tr_{\parallel} \log(-\d^2_{\parallel} + m^2)_1) = \frac{1}{24} \left(n-\frac{1}{n}\right) \log(m^2)\eeq
Thus,
\beq \log\frac{Z_n}{Z^n} = \frac{N}{24} \left(n-\frac{1}{n}\right) \sum_{\vec{k}_{\perp}} \log(\vec{k}^2_{\perp})
= - N \frac{ \pi}{6} \left(n-\frac{1}{n}\right) L^{D-2} G^L_1(x,x) \label{LGauss1}\eeq
Here, $G^L_n(x,x)$ is the free propagator on an $n$-sheeted Riemann surface, which incorporates the finite size effects in the transverse direction. Explicitly,
\beq G^L_n(x,x') = \frac{1}{L^{D-2}} \sum_{\vec{k}_{\perp}} G^{D=2}_n(x_{\prl},x'_{\prl}; k^2_{\perp}) e^{i \vec{k}_{\perp} (\vec{x}_{\perp} - \vec{x}'_{\perp})} \label{GLn}\eeq
In particular, for  $n =1$,
\beq G^L_1(x,x') = \frac{1}{L^{D-2}} \sum_{\vec{k}_{\perp}} \int \frac{d^2 k_{\prl}}{(2 \pi)^2} \frac{1}{k^2_{\prl} + k^2_{\perp}} e^{i k (x-x')}\eeq
justifying the last step in eq. (\ref{LGauss1}).

An alternative representation for the propagator (\ref{GLn}) on the torus can be obtained by Poisson resumming $\vec{k}_{\perp}$, which is equivalent to ``periodizing" the infinite volume propagator,
\beq G^L_n(x,x') = \sum_{\vec{l}} e^{i \vec{l}\vec{\varphi}} G_n(x+ \vec{l} L, x')\label{periodize}\eeq
where $\vec{l}$ is a vector of $D-2$ integers in the plane parallel to the boundary. Note that when $x =x'$, only the $l = 0$ term in eq. (\ref{periodize}) is ultra-violet divergent and $G^L_n(x,x) - G_n(x,x)$ is finite. Moreover, since the $l = 0$ term, $G_1(x,x) \sim \Lambda^{D-2}$, is $L$ independent, it gives a non-universal contribution to $\log (Z_n/Z^n)$, eq. (\ref{LGauss1}), proportional to the area of the boundary. Concentrating on the universal constant term,
\beq\log\frac{Z_n}{Z^n} = - N \frac{ \pi}{6} \left(n-\frac{1}{n}\right) L^{D-2} (G^L_1(x,x) - G_1(x,x))\label{freeL}\eeq
where from eqs. (\ref{G1D}), (\ref{periodize}),
\beq L^{D-2} (G^L_1(0) - G_1(0)) = \frac{\Gamma(D/2-1)}{4 \pi^{D/2}}\sum_{\vec{l} \neq 0} \frac{e^{i \vec{l} \vec{\varphi}}}{|\vec{l}|^{D-2}}\eeq
Here and below we abbreviate $G^L_1(x,x)$ by $G^L_1(0)$. 


We can now explicitly evaluate the universal constant contribution $\gamma_n$ to the entanglement entropy for $D = 3$ and $D = 4$. 
\bea 
\gamma_n &=& -\frac{N}{12} \left(1 + \frac{1}{n}\right) \log(2 |\sin \varphi/2|), \quad\quad D = 3 \\
\gamma &=& - \frac{N}{6} \log(2 |\sin \varphi/2|), \quad\quad D = 3\eea

For $D = 4$, we note that the sum 
\beq \sum_{\vec{l} \neq 0} \frac{e^{i \vec{l} \vec{\varphi}}}{\vec{l}^{2}} = (2 \pi)^2 G^{D=2}(\vec{\varphi})\eeq
where $G^{D=2}(\vec{\varphi})$ is the massless two-dimensional propagator (with the zero-mode removed) on a torus with side-length $2\pi$. This propagator can be expressed in terms of the Jacobi-theta function $\theta_1$,
\beq G^{D=2}(\vec{\varphi}) = - \frac{1}{2 \pi} \left( \log\Big|\theta_1\big(\frac{\varphi_1+i \varphi_2}{2 \pi}, i\big)\Big| - \frac{\varphi^2_2}{4 \pi} - \log \eta(i)\right)\eeq
where $\eta$ is the Dedekind-eta function.

Thus,
\bea \gamma_n &=& \frac{\pi N}{6} \left(1+\frac{1}{n}\right) G^{D=2}(\vec{\varphi})
, \quad\quad D = 4\\
\gamma &=& \frac{\pi N}{3} G^{D=2}(\vec{\varphi}) 
, \quad \quad D = 4 \eea

\subsection{$4-\epsilon$ expansion}
We now compute the universal finite size correction to leading order in $4-\epsilon$ expansion. 
The leading correction to the free theory behaviour comes from the boundary perturbation (\ref{boundary}), as at the fixed point $c^*_r \sim \sqrt{\epsilon}$ for $1 -n \gg \epsilon$ and $c^*_r \sim (n-1)$ for $|1-n| \ll \epsilon$. Thus, 
\beq \delta^{1,0} \log\frac{Z_n}{Z^n} = -\frac{c_r}{2} \int d^{D-2} x_{\perp} \langle \phi^2(r = 0)\rangle_n = -\frac{N c_r}{2} L^{D-2} G^L_n(r=r'=0) = - \frac{N c_r}{2 n} L^{D-2} G^L_1(x,x) \label{boundaryL}\eeq
where in the last step we've used eqs. (\ref{GnG1}), (\ref{periodize}).
Again, subtracting the non-universal area law piece $\sim L^{D-2} G_1(0)$, and combining eq. (\ref{boundaryL}) with the free theory result (\ref{freeL}),
\beq \log\frac{Z_n}{Z^n} = - N \left(\frac{\pi}{6} \left(n - \frac{1}{n}\right) + \frac{c_r}{2n}\right) L^{D-2} (G^L_1(0) - G_1(0))\label{ZLO1}\eeq
Now replacing $c_r$ by it's fixed point value and taking $D \to 4$,
\beq \gamma_n = N \left(\frac{\pi}{6} \left(1 + \frac{1}{n}\right) + \frac{c^*_r}{ 2 n (n-1)}\right) G^{D=2}(\varphi,\varphi)
\eeq
Here we've set all the twists $\varphi_i$ equal. For $1- n \gg \epsilon$, eq. (\ref{cfar1}), the $c^*$ term  gives a correction of order $\sqrt{\epsilon}$ to the free theory result. However, in the limit $|1 -n| \ll \epsilon$, eq. (\ref{cclose1}),
the correction due to the boundary perturbation cancels with the free theory result to leading order in $n - 1$, leaving,
\beq \gamma_n \stackrel{n \to 1}{\approx} -\frac{\pi N (N+8)}{9 (N+2)} \frac{n-1}{\epsilon} G^{D=2}(\varphi,\varphi)
\eeq
This implies that at the Wilson-Fisher fixed point the universal finite size correction to the entanglement entropy,
\beq \gamma \sim O(\epsilon)\eeq
parametrically smaller than at the Gaussian fixed point in $D = 4- \epsilon$. 

\subsection{Beyond the leading order in $\epsilon$}
\begin{figure}[t]
\begin{center}
\includegraphics[width=4in]{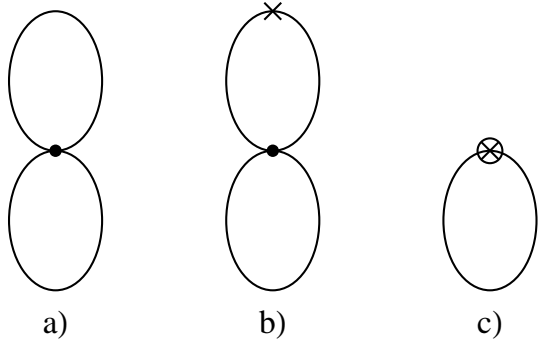}
\caption{Contributions to the partition function at order $u$.}
\label{FigZu}
\end{center}
\end{figure}
We now evaluate the universal finite size correction to the entanglement entropy $\gamma$ to order $\epsilon$. As before, we only work to leading order in $n-1$. To order $u$ the partition function receives contributions from the diagrams in Fig. \ref{FigZu}. The diagram in Fig. \ref{FigZu} a) is given by,
\bea \delta^{0,1}\log\frac{Z_n}{Z^n} &=& - \frac{N(N+2) u_r \mu^{\epsilon} }{4} \int d^D x\, \left(G^L_n(x,x) - G^L_1(x,x)\right) \times \nn\\
&& \big[(G^L_n(x,x) - G_1(x,x)) + (G^L_1(x,x) - G_1(x,x))\big]\nn\\
&\stackrel{n \to 1}\approx& -\frac{N(N+2) u_r \mu^{\epsilon}}{2} (G^L_1(0) - G_1(0)) \sum_{\vec{k}_{\perp}} \int d^2 x\, \left(G^{D=2}_n(x,x;k^2_{\perp}) - G^{D=2}_1(x,x;k^2_{\perp})\right) \nn\\
&=& \frac{N(N+2) (n-1) u_r \mu^{\epsilon} }{12} (G^L_1(0) - G_1(0)) \sum_{\vec{k}_{\perp}} \frac{1}{\vec{k}^2_{\perp}}\eea
where in the last step we've used eq. (\ref{CardyD2int}).

The diagram in Fig. \ref{FigZu} b) can be evaluated with $n=1$ propagators,
\bea \delta^{1,1}\log\frac{Z_n}{Z^n} &=& \frac{N (N+2) u_r c_r \mu^{\epsilon}}{2} (G^L_1(0) - G_1(0)) \int d^{D-2} x_{\perp} \int d^{D}x'\, G^L_1(x_\perp,x')^2 \nn\\
&=& \frac{N(N+2) u_r c_r \mu^{\epsilon}}{8 \pi} (G^L_1(0) - G_1(0)) \sum_{\vec{k}_{\perp}} \frac{1}{k^2_{\perp}}\eea

Finally, the diagram in Fig. \ref{FigZu} c) can be obtained from eq. (\ref{boundaryL}) by substituting the counterterm for $c$, eq. (\ref{delta1c}). Combining all the diagrams in Fig. \ref{FigZu} with the $O(1)$ result, eq. (\ref{ZLO1}),
\bea \log \frac{Z_n}{Z^n} &=& - \frac{N}{2} \left(\frac{2 \pi}{3} (n-1) + c_r\right) L^{D-2} (G^L_1(0) - G_1(0)) 
\nn \\
&\times & \left[ 1- \frac{(N+2) u_r}{4 \pi} \left( (\mu L)^{\epsilon} \sum_{\vec{k}_{\perp}} \frac{1}{(L k_{\perp})^2} - \frac{1}{2 \pi \epsilon}\right)\right]\label{ZnOu}\eea
Applying the usual technique for analytically continuing sums over $D$-dimensional vectors,
\beq \sum_{\vec{k}_{\perp}} \frac{1}{(L k_{\perp})^2} = \int_0^{\infty} ds \, T(s)^{D-2}\label{sumkperp}\eeq
where
\beq T(s) = \sum_n e^{-s (2 \pi n + \varphi)^2} \label{Tdef}\eeq
The function $T(s)$ has the following asymptotics,
\bea T(s) &\to&  \frac{1}{\sqrt{4 \pi s}}, \quad s \to 0 \label{Tass1}\\
T(s) &\to& e^{-s \varphi^2}, \quad s \to \label{Tass2}\infty
\eea
Hence, for finite $\varphi$ the integral in eq. (\ref{sumkperp}) converges in the $s \to \infty$ region. Moreover, the $s \to 0$ region contributes a pole for $D \to 4$,
\beq \sum_{\vec{k}_{\perp}} \frac{1}{(L k_{\perp})^2} \to  \frac{1}{2 \pi \epsilon} + \mbox{finite terms}\eeq
As expected, this pole precisely cancels with the $c$ counterterms, so that the expression (\ref{ZnOu}) is finite. Moreover, setting $c_r$ to its fixed point value, eq. (\ref{cstar}), the prefactor in eq. (\ref{ZnOu}) is already $O(\epsilon)$, so that we can neglect the $O(u)$ terms in the square brackets. Thus,
\beq \log\frac{Z_n}{Z^n} = - \frac{N \pi \epsilon (n-1)}{3(N+8)} L^{D-2} (G^L_1(0) - G_1(0))\eeq
and
\beq \gamma = \frac{N \pi \epsilon}{3 (N+8)} G^{D=2}(\varphi,\varphi)
\label{gammafinal}\eeq
Note that the result (\ref{gammafinal}) is of $O(1)$ in $N$ for $N \to \infty$, instead of the naively expected $O(N)$. It is not clear if this is an artifact of working to leading order in $\epsilon$.
\begin{figure}[t]
\begin{center}
\includegraphics[width=4in]{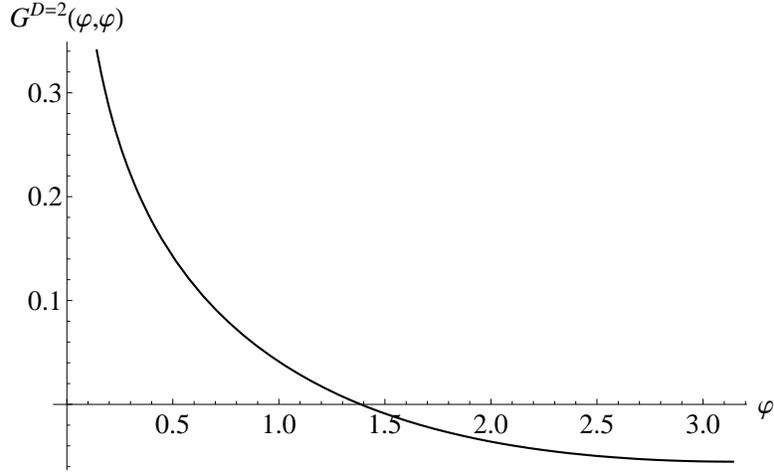}
\caption{The function $G^{D=2}(\varphi,\varphi)$ determining the dependence of $\gamma$ on the twist $\varphi$, eq. (\ref{gammafinal}).}
\label{FigGphi}
\end{center}
\end{figure}

The function $G^{D=2}(\varphi,\varphi)$ which determines the $\varphi$ dependence of $\gamma$ is shown in Fig. \ref{FigGphi}. We observe that $\gamma$ is a monotonically decreasing function of $\varphi$ for $0 < \varphi < \pi$. In particular, for $\varphi = \pi$,
\beq \gamma = - \frac{N \epsilon}{12 (N+8)}\log 2\eeq
Thus, $\gamma$ is negative for anti-periodic boundary conditions.  On the other hand, for $\varphi \to 0$,
\beq \gamma {\approx} - \frac{N \epsilon}{6 (N+8)} \log \varphi, \quad \varphi \to 0\label{philog}\eeq
suggesting that $\gamma$ is positive for periodic boundary conditions. Note that our expression for $\gamma$ becomes invalid for $\varphi$ sufficiently 
small. The value of $\varphi$ where the breakdown of direct perturbative expansion occurs can be estimated as follows. Let us separate out the quasi-zero mode $\phi_0$ of the field $\phi$,
\beq \phi(x) = \frac{1}{L^{(D-2)/2}} \phi_0(x_{\prl}) e^{i \vec{\varphi} \cdot \vec{x}_{\perp}/L} + \tilde{\phi}(x) \eeq  where $\tilde{\phi}(x)$ has
the $\vec{k}_{\perp} = \frac{\vec{\varphi}}{L}$ mode omitted. At the mean-field level, the effective action for $\phi_0$ is a two dimensional $\phi^4$ field theory, with an effective mass $m^2_{2D} \sim \frac{\varphi^2}{L^2}$ and quartic coupling $u_{2D} \sim \frac{u}{L^{D-2}}$. We know that perturbative expansion in a $2D$ theory is valid for $u_{2D}/m^2_{2D} \ll 1$. Thus, setting $D = 4$ and $u = u^*$, we obtain,
\beq \varphi^2 \gg \epsilon\eeq
as the domain of validity of perturbation theory. For smaller values of $\varphi$, the zero mode must be treated separately and non-perturbatively. This result can be checked in the $1/N$ expansion where one obtains a slightly stronger condition $\varphi^2 \gg \epsilon \log \varphi$. Cutting off the logarithmic divergence of (\ref{philog}) at the value of $\varphi$ where perturbation theory breaks down, we obtain,
\beq \gamma \approx - \frac{N \epsilon}{12 (N+8)} \log \epsilon \label{twist0}\eeq
We conjecture that eq. (\ref{twist0}) is the leading order result for the case of zero twist (periodic boundary conditions).

\section{Large $N$ limit}\label{sec:largeN}
In this section we compute the correlation length correction to the Renyi entropy $S_n$, eq. (\ref{sndef}), in the large $N$ limit. Although we are mainly interested in the physical case $D = 3$, we will keep the dimension of space-time arbitrary in our discussion in order to compare the results of the large-$N$ and $4-\epsilon$ expansions. 

When working in the large-$N$ limit, it is more convenient to use the non-linear $\sigma$-model version of the $O(N)$ model (\ref{soft}), where the quartic interaction is replaced by a local constraint $\phi^2(x) = \frac{1}{g}$. Enforcing this constraint with the help of the Lagrange multiplier $\lambda(x)$, the action takes the form,
\beq S = \int d^Dx \left(\frac{1}{2} (\d_{\mu} \phi)^2 + \frac{1}{2} i \lambda (\phi^2 - \frac{1}{g})\right) \label{nonlinear}\eeq
Our discussion in section \ref{sec:general} is then directly transcribed into the present case with the replacement, $t \to -(\frac{1}{g} - \frac{1}{g_c})$, $\phi^2 \to i \lambda$.  In particular, to determine the coefficient $r_n$ of the correlation length correction to leading order in $1/N$, we need to find the behaviour of $\langle i \lambda(x)\rangle$ at the critical point. 

We tune the $O(N)$ model to criticality $g = g_c$. At $N = \infty$, the problem is reduced to finding the saddle-point value of the Lagrange multiplier $\langle i \lambda(x)\rangle_n$ such that the gap equation,
\beq G_n(x,x) = \frac{1}{N} \langle \phi^2(x) \rangle_n  = \frac{1}{N g_c}\eeq
is satisfied. Here $G_n(x,x')$ is the Green's function of the operator $-\d^2 + \langle i \lambda(x) \rangle_n$ on the $n$-sheeted Riemann surface. The quantity $G_n(x,x)$ requires regularization; we will implicitly use point splitting regularization. It is convenient to rewrite the gap equation as,
\beq G_n(x,x) - G_1(x,x) = 0 \label{gap2}\eeq

We note that at $N = \infty$ the scaling dimension of $\lambda(x)$ is $2$, so,
\beq \langle i \lambda(x) \rangle_n = \frac{a_n}{r^2}\label{lambda}\eeq
From (\ref{tphi2}), with the appropriate replacement $\phi^2 \to i \lambda$, $t \to g^{-1}_c - g^{-1}$, the constant $a_n$ is related to the constant $d_n$ (\ref{fshort}) as
\beq d_n = \frac{1}{m^{D-2}} \left(\frac{1}{g_c} - \frac{1}{g}\right) a_n \label{darel}\eeq
Now from the gap equation at finite $m$,
\beq  \frac{1}{N g} - \frac{1}{N g_c} = \int \frac{d^D p}{(2 \pi)^D} \left(\frac{1}{p^2 + m^2} - \frac{1}{p^2}\right) = \frac{1}{(4\pi)^{D/2}} \Gamma(1-D/2) m^{D-2}  \label{mN}\eeq
and
\beq d_n = -\frac{N}{(4\pi)^{D/2}} \Gamma(1-D/2) \,a_n \label{dacon}\eeq
In particular in $D = 3$, $d_n = \frac{N}{4 \pi} a_n$. Thus, the problem of computing the entanglement entropy at $N = \infty$ reduces to finding the constants $a_n$. 

We now need to find the Green's function $G_n$. The main observation is that the angular harmonics on an $n$-sheeted Riemann surface are $\frac{1}{\sqrt{2 \pi n}} e^{i l \theta/n}$, where $l$ is an integer. Hence,
\beq G_n(x,x') = \int \frac{d^{D-2}k_{\perp}}{(2\pi)^{D-2}} e^{i k_{\perp} (x_\perp - x'_\perp)} G^{D=2}_n(r,r',\theta; k^2_{\perp})\label{perpint}\eeq
where the two-dimensional massive propagator on an $n$-sheeted Riemann surface is given by,
\beq G^{D=2}_n(r,r',\theta; m^2) = \sum_{l} \frac{e^{i l (\theta-\theta')/n}}{2 \pi n} g_{l}(r,r';m^2) \label{spectral} \eeq
Here, 
\beq \left(-\frac{1}{r} \frac{\d}{\d r} \left(r \frac{\d}{\d r}\right) + \frac{(l/n)^2 + a_n}{r^2} + m^2 \right) g_{l}(r,r';m^2) = \frac{1}{r} \delta(r-r')\eeq
We use spectral decomposition for $g_{l}$,
\beq g_{l}(r,r';m^2) = \int dE \frac{1}{E + m^2} \phi_{l,E}(r) \phi^*_{l,E}(r')\eeq
where
\beq \left(-\frac{1}{r} \frac{\d}{\d r} \left(r \frac{\d}{\d r}\right) + \frac{(l/n)^2 + a_n}{r^2}\right) \phi_{l, E} = E \phi_{l, E} \label{SE}\eeq
and $\phi_{l,E}$ are normalized to
\beq \int dr r \phi^*_{l, E}(r) \phi_{l, E'}(r) = \delta(E-E')\eeq
The constant $a_n$ must be positive in order to avoid the presence of negative energy states, which would render our saddle point unstable. Let us call the quantity $l^2/n^2 + a_n = \nu^2$. Eq. (\ref{SE}) admits two linearly independent solutions, 
\bea \phi(r) &=& \frac{1}{\sqrt{2}}J_{|\nu|}(\sqrt{E} r)\label{plus}\\ \phi(r) &=& \frac{1}{\sqrt{2}} J_{-|\nu|}(\sqrt{E} r)\label{minus}\eea
 We recall that
\beq J_{\nu}(x) \sim |x|^{\nu}, \quad x \to 0 \label{xnu}\eeq
When working in free space (in the absence of conical singularity and potential (\ref{lambda})) one chooses only the solutions with a positive index $|\nu| = |l|$, so that $\phi_E(r)$ is finite and differentiable at $r = 0$. However, in the present problem there is no {\it a priori} physical reason why the solutions (and hence the propagator) have to remain finite as $r \to 0$. 

In fact, a particle in a $1/r^2$ potential is a famous problem known as conformal quantum mechanics.  Note that the potential (\ref{lambda}) is highly singular and requires regularization at short distances. Such regularization will automatically appear in the linear $O(N)$ model, which can be obtained from (\ref{nonlinear}) by adding a term $\lambda^2/4u$ to the Lagrangian. In that case, eq. (\ref{lambda}) only holds for $u r^{4-D} \gg 1$ and the saddle point value $\langle i \lambda \rangle$ is modified at short distances. We note that even after this regularization, the $l \neq 0$ states still experience an $l^2/r^2$ centrifugal barrier and we must choose positive index solutions (\ref{plus}) for them. We now concentrate on the $l = 0$ sector. For simplicity, imagine cutting the $1/r^2$  divergence off at some radius $r=r_0$ and replacing it by a finite potential. 
Generally, the resulting scattering states will approach the positive index solutions (\ref{plus}) for $\sqrt{E} r_0 \to 0$. However, non-trivial behaviour can occur if the potential is close to developing a bound state. In that case, for $|\nu| < 1$, one ``dynamically" generates a length-scale $\xi$ and the scattering solutions become linear combinations of (\ref{plus}) and (\ref{minus}) with coefficients (and, thus, the phase-shifts) depending on $\sqrt{E} \xi$. Since we are looking for a scale invariant solution to the gap equation, we need $\xi \to \infty$, i.e. the system is exactly at the threshold of bound state formation. At this threshold, for $\sqrt{E} r_0 \to 0$ one obtains negative index solutions (\ref{minus}). Note, that this behaviour is special to the range $|\nu| < 1$ and does not occur for $|\nu| > 1$. This fact could be anticipated as the negative index solutions are square integrable at short distances for $|\nu| < 1$ but not for $|\nu| > 1$. 

Thus, applying RG terminology to the simple quantum mechanics problem (\ref{SE}), we conclude that there are two fixed points - one stable (\ref{plus}) and one unstable (\ref{minus}). However, we are allowed to choose the unstable fixed point solutions as we are fine tuning both the long and short distance parts of $\langle i \lambda \rangle$ to solve the gap equation.

With these remarks in mind, 
\bea g_{l}(r,r';m^2) &=& \int_0^{\infty} k dk \frac{1}{k^2 + m^2} J_{\nu_l}(k r) J_{\nu_l} (k r')\label{BesselInt}
\eea
where $\nu_l = \alpha$ for $l = 0$ and $\nu_l = \sqrt{l^2/n^2+ \alpha^2}$ for $|l| > 0$, with $a_n = \alpha^2$. The constant $\alpha$ can be either positive or negative. We note that as discussed in Ref \onlinecite{VBSVortex}, $\alpha$ enters the operator product expansion of the field $\phi(x)$ as $x$ approaches the conical singularity,
\beq \phi(x_{\prl},x_{\perp}) \sim r^{\alpha} \phi(0,x_{\perp}), \quad r \to 0 \label{OPE}\eeq

Combining eqs. (\ref{perpint}),(\ref{spectral}) and (\ref{BesselInt}), and performing the integrals over $k_{\perp}$, $k$ we obtain
\beq G_n(r =r',\theta, x_{\perp} = x'_{\perp}) = \frac{\Gamma((3-D)/2)}{2 \pi n  (4 \pi)^{(D-1)/2} r^{D-2}} \sum_l \frac{\Gamma(D/2-1 + \nu_l)}{\Gamma(2-D/2+\nu_l)} e^{i l \theta/n} \label{sumlGamma}\eeq
Since we are mostly interested in $G_n(x=x')$, we have set $r = r'$, $x_{\perp} = x'_{\perp}$ in (\ref{sumlGamma}); we have left $\theta \neq 0$ as a regulator. 

As an aside that will be of some interest later, we note that (\ref{sumlGamma}) is meaningful only for $\alpha > -(D/2-1)$. For $\alpha \leq -(D/2-1)$ one obtains an infrared divergence in the $k_{\perp}, k \to 0$ region of integrals (\ref{perpint}),(\ref{BesselInt}). We note that at $\alpha = -(D/2-1)$, eq. (\ref{SE}) has a zero energy solution,
\beq \phi(r) = \frac{1}{r^{D/2-1}}\label{phicond}\eeq
The solution (\ref{phicond}) could, in principle, correspond to a saddle point with a non-zero expectation value  $\langle \phi(x)\rangle$. Note that the $r$ dependence of (\ref{phicond}) is consistent with the scaling dimension $[\phi(x)] = D/2-1$ in the $N \to \infty$ limit. Alternatively, observe that the scaling dimension of the ``boundary" operator, $[\phi(0,x_{\perp})] = D/2-1 + \alpha \to 0$ as $\alpha \to -(D/2-1)$, indicating a tendency to condense. However, the infrared divergence of the propagator (\ref{sumlGamma}) indicates that condensation of $\phi(x)$ at the conical singularity is unstable to fluctuations. This is not unexpected, as the condensate would be $D- 2 < 2$ dimensional. Such a condensate certainly cannot exist for any $g > g_c$, as it would violate the Mermin-Wagner theorem. Long range interactions could potentially stabilize the condensate exactly at the critical point, however, the above discussion shows that this does not occur (at least in the large-$N$ limit).

We use contour integration to write (\ref{sumlGamma}) in a somewhat more convenient form,
\bea && G_n(r=r',\theta,x_\perp = x'_\perp) \nn \\ 
&&= \frac{1}{4 \pi^{D/2} \Gamma(2-D/2) r^{D-2}} \Big(\int_0^{\infty} d\nu \frac{\nu}{\sqrt{\nu^2+\alpha^2}} U_{\sqrt{\nu^2+\alpha^2}}(\theta) \,R(\nu) + \theta(-\alpha) \frac{i R(i \alpha)}{n}\Big)\nn\\\label{GR}\eea
with
\bea 
U_{\nu}(\theta) &=& \frac{\cosh(\nu(\pi n - |\theta|))}{\sinh(\pi n \nu)}\label{Un}\\
R(\nu) &=& -i\Gamma(3-D)
\left[\frac{\Gamma(-i \nu+D/2-1)}{\Gamma(-i \nu + 2-D/2)} - \frac{\Gamma(i \nu+D/2-1)}{\Gamma(i \nu + 2-D/2)}\right]\label{Rnu}\\ &=& \frac{2 \pi \Gamma(3-D) \sin(\pi(3-D)/2) \sinh(\pi \nu)}{\cosh^2 \pi \nu - \sin^2(\pi(3-D)/2)} \frac{1}{|\Gamma(i\nu + 2-D/2)|^2}\eea
In particular, for $D = 3$, $R(\nu) = \pi \tanh(\pi \nu)$. We note that despite the presence of the $\theta(-\alpha)$ term in eq. (\ref{GR}), $G_n(r=r',\theta,x_{\perp} = x'_{\perp})$ is  analytic at $\alpha = 0$ as is evident from eq. (\ref{sumlGamma}).
Thus, the gap equation (\ref{gap2}) takes the form,
\bea 
&& (G_n-G_1)\big|_{x=x'} \nn \\ &&= \frac{1}{4 \pi^{D/2} \Gamma(2-D/2) r^{D-2}} \Bigg[\int_0^{\infty} d\nu   \left(\frac{\nu}{\sqrt{\nu^2+\alpha^2}} \coth(\pi n \sqrt{\nu^2+\alpha^2}) - \coth(\pi \nu)\right) R(\nu)\nn \\ &&~~~~~~+ \theta(-\alpha) \frac{i}{n} R(i \alpha)\Bigg]= 0\label{gapR}\eea
The function $R(\nu)$ is positive for real values of $\nu$. So the left-handside of the gap equation goes to $- \infty$ as $\alpha \to \infty$ and to $\infty$ as $\alpha \to - (D/2-1)^+$. Hence, the gap equation always has at least one solution, and more generally, an odd number of solutions. Numerically, we find that the gap equation has a unique solution for all $n$ for $D < D_c$, $D_c \approx 3.74$. For $D > D_c$, there are one or three solutions depending on the value of $n$, as we will discuss below.

As we are mainly interested in the entanglement entropy, let us consider the limit $n \to 1$. Then we expect $\alpha \to 0$. The integral in (\ref{gapR}) is non-analytic at $\alpha = 0$, due to singular behaviour in the $\nu \to 0$ region. Noting that $R(\nu) \approx R'(0) \nu$, as $\nu \to 0$, we obtain to leading order in $\alpha$,
\bea &&(G_n-G_1)\big|_{x=x'}-(G_n-G_1)\big|_{x=x', \alpha = 0}\nn\\ &\approx& \frac{R'(0)}{4 n \pi^{D/2} \Gamma(2-D/2) r^{D-2}}\left(\frac{1}{\pi}\int_0^{\infty} d\nu \left(\frac{\nu^2}{\nu^2 + \alpha^2} -1\right) - \theta(-\alpha) \alpha\right)\nn\\
&=& \frac{R'(0)}{4 n \pi^{D/2} \Gamma(2-D/2) r^{D-2}}\left(-\frac{1}{2}|\alpha| - \theta(-\alpha) \alpha\right) = - \frac{\Gamma(D/2-1)^2 \Gamma(D/2)}{4 \pi^{D/2} \Gamma(D-1) r^{D-2}}\frac{\alpha}{n}\nn \\\label{alpha0} \eea
where the contributions from the integral and the $\theta$ function have combined to produce a result analytic in $\alpha$. Now using eqs. (\ref{GJ}), (\ref{JD}) for $(G_n-G_1)\big|_{x=x', \alpha = 0}$,
\beq \alpha \approx - \frac{D-2}{2(D-1)} (n-1), \quad n \to 1 \label{alphan1}\eeq
Note that the exponent $\alpha$ controlling the OPE (\ref{OPE}) of the field $\phi(x)$ at the conical singularity  is positive for $n < 1$ and negative for $n > 1$. Now, from (\ref{alphan1}),
\beq a_n = \frac{(D-2)^2}{4(D-1)^2} (n-1)^2, \quad n \to 1\eeq
Therefore, combining eqs. (\ref{renh}) and (\ref{darel}), we find that
\beq r_n \propto \frac{a_n}{1-n} \propto n-1, \quad n \to 1\eeq
and the correlation length correction to the entanglement entropy proper vanishes at leading order in $N$, 
\beq r = \lim_{n \to 1} r_n = 0\eeq
Thus, for all dimensions $2 < D< 4$
\beq r \sim O(N)\eeq
even though $r_n \sim O(N^2)$ for all $n \neq 1$. 

So far we have concentrated on the solution to the gap equation in the $n \to 1$ limit for arbitrary dimension. However, we can also obtain an analytic solution for arbitrary $n$ in the limit $D = 4-\epsilon$. Such a solution is useful for comparison to the results of the $4-\epsilon$ expansion presented in section \ref{sec:4eps}.

When $D = 4-\epsilon$, the function $R(\nu) = - 2 \nu^{1-\epsilon} \Gamma(-1+\epsilon)$. The divergence of the $\Gamma$ function is not important here as it is just an overall factor in the gap equation (which anyway cancels with $\Gamma(2-D/2)$ in (\ref{GR})). However, the integral (\ref{gapR}) now diverges for $\nu \to \infty$ if $\epsilon = 0$. Hence, for generic $n$ and $D = 4-\epsilon$ the leading $\alpha$-dependent contribution to the gap equation comes from the region $\nu \gg 1$ and is of order, $\frac{1}{\epsilon} \alpha^2$. This suggest that $\alpha$ will be at most of order $\epsilon^{1/2}$. However, for $\alpha$ very small (i.e. $n \to 1$), we already know from the previous discussion that the leading contribution to the integral scales as $|\alpha|$ and comes from the $\nu \to 0$ region. Keeping these two contributions (one non-analytic in $\alpha$ and the other analytic, but with a diverging coefficient) and setting $\alpha = 0$ in the rest of the integral, we reduce the gap equation to
\bea
&& \frac{1}{\pi n} \int_0^{\infty} d\nu \left(\frac{\nu^2}{\nu^2+\alpha^2}-1\right) + \int_0^{\infty} d\nu \, \nu \left(\frac{\cosh(\pi n \nu)}{\sinh(\pi n \nu)} - \frac{\cosh(\pi  \nu)}{\sinh(\pi  \nu)}\right)\nn \\
&&~~~~~- \frac{1}{2} \alpha^2 \int_{\nu \gg 1}^{\infty} d \nu \, \nu^{-\epsilon-1} - \theta(-\alpha)\frac{\alpha}{n} =0 \label{gap4eps}\eea
\beq \frac{\alpha}{n} + \frac{\alpha^2}{\epsilon} - \frac{1}{6} \left(\frac{1}{n^2} - 1\right) =0 \label{gapa}\eeq
The quadratic has two solutions,
\beq \alpha_{\pm} = - \frac{\epsilon}{2n} \pm \frac{1}{2 n} \sqrt{\epsilon^2 + \frac{2 \epsilon}{3}(1-n^2)}\label{alphapm}\eeq
and the corresponding values of $d_n$, eq. (\ref{dacon}), are,
\beq d^{\pm}_n = \frac{N}{8 \pi^2} \left[ \frac{1}{6}\left(\frac{1}{n^2}-1\right) + \frac{\epsilon \mp \sqrt{\epsilon^2 + \frac{2}{3} \epsilon (1-n^2)}}{2 n^2}\right] \label{largeNcross}\eeq
Eq.  (\ref{largeNcross}) is in agreement with the result of the $4-\epsilon$ expansion, eq. (\ref{dn4eps}), and we can identify the $\alpha_\pm$  saddle points with the $c^\pm_r$ fixed points. Moreover, we see that the predictions of the large-$N$ (\ref{alphapm}) and $4-\epsilon$ expansion (\ref{alpha4eps}) for the OPE exponent $\alpha$ also agree.
Note that both saddle points (\ref{alphapm}) disappear for $n > n_c \approx 1 + 3 \epsilon/4$. This coincides with the value of $n$ at which runaway of RG flow is observed in the $4-\epsilon$ expansion. However, as we noted earlier, the gap equation always has an odd number of solutions. Thus, we have missed a solution in our discussion above. This solution has $\alpha \approx -(D/2-1) \to -1$, i.e. $\alpha$ is not small. Its existence is possible due to a cancellation of $1/\epsilon$ divergences between the large $\nu$ part of the integral and the $\theta(-\alpha)$ term in (\ref{gapR}). Keeping these two contributions to the gap equation, we obtain in the $\alpha \to -(D/2-1)$ limit,
\beq \frac{\alpha^2}{\epsilon} - \frac{\epsilon}{n}\frac{1}{\alpha + D/2-1} = 0\eeq
So,
\beq \alpha = -1 + \frac{1}{2} \epsilon + \frac{1}{n} \epsilon^2 \label{ncond}\eeq
Eqs. (\ref{alphapm}), (\ref{ncond}) comprise the three solutions to the gap equation for $1 < n < n_c$, and eq. (\ref{ncond}) is the only solution for $n > n_c$. We speculate that the runaway of the RG flow observed in $4-\epsilon$ expansion for $n > n_c$ is towards the fixed point (\ref{ncond}). As we noted above, the value $\alpha = -(D/2-1)$ corresponds to the would be condensation of the $\phi$ field at the conical singularity. Thus, for $\epsilon \to 0$, the saddle-point (\ref{ncond}) is proximate to such condensation. This is consistent with our interpretation of the RG flow $c \to -\infty$ as the tendency to formation of $\langle \phi(x) \rangle \neq 0$. However, the large-$N$ analysis demonstrates that no true spontaneous symmetry breaking at the conical singularity occurs for $D < 4$. 

To our knowledge no such non-trivial $n$-dependence has been previously observed in any theories.  Still, in the large-$N$ expansion such behaviour is only present for $D > D_c \approx 3.74$ and its relevance to the physical case $D = 3$ is doubtful. Moreover, the non-analyticity occurs away from $n = 1$ and, thus, is unimportant for computing the entanglement entropy proper. Indeed, the behaviour of the theory for $n \to 1$ (\ref{alphan1}) is found to evolve smoothly as the dimension $D$ increases from $2$ to $4$. 




\begin{table}
\medskip
\begin{center}
\begin{tabular}{|c|c|}
\hline
$n$ & $\alpha_n$\\
\hline
2 & -0.16515\\
3 & -0.26594\\
4 & -0.32905\\
5 & -0.36743\\
\hline
\end{tabular}
\end{center}
\caption{Solution to the gap equation in the large-$N$ limit for $D = 3$.}\label{Tblalpha}
\end{table}

We now come back to the physical case $D = 3$, where the solution to the gap equation is unique. The numerical solution for the first few integers $n$ is listed in Table \ref{Tblalpha}. Then, from (\ref{finalZn}) and (\ref{darel}),
\beq r_n =   \frac{3 \pi^2 N^2}{128} \frac{n \alpha^2_n}{n-1} , \quad D = 3\label{trrhonD3}\eeq
The coefficient (\ref{trrhonD3}) can be, in principle, obtained numerically by performing classical Monte-Carlo simulations of the $O(N)$ model in the spirit of Ref. \onlinecite{Buivid}.

So far our large-$N$ computation has been confined to the correlation length correction to the Renyi entropy. At leading order the calculation was technically fairly simple, as utilizing the discussion in section \ref{sec:general}, we could work at the critical point. In particular, the form of the Lagrange multiplier $\langle i \lambda(r)\rangle$ was fixed by scale invariance up to an overall constant. To proceed beyond the leading order, as is required for the calculation of the correlation length correction to the entanglement entropy proper, we would have to work in the gapped phase. The Lagrange multiplier $\langle i \lambda(r) \rangle$ would now be a non-trivial function of $r$ with a length scale determined by the correlation length $\xi = m^{-1}$. Similarly, if we wish to compute the finite size correction $\gamma$ to the entanglement entropy, $\langle i \lambda(r) \rangle$ will again vary non-trivially with a length scale determined by the size $L$ of the compact direction. In both cases, we have to solve the gap equation for a whole function $\langle i \lambda(r) \rangle$ rather than a single number $a_n$. In principle, this problem can be addressed numerically. It would be particularly interesting to check whether $\gamma \sim O(1)$ for $N \to \infty$ as suggested by the $4-\epsilon$ expansion, eq. (\ref{gammafinal}).



\section{Conclusion. Future Directions.}\label{sec:concl}
In the present work we have computed the universal finite size and correlation length corrections to the entanglement entropy and the Renyi entropy for the $O(N)$ model. The evaluation of this entropy required a study of the $O(N)$ field theory on a $n$-sheeted Riemann surface
for general $n$, and an understanding of the nature of the $n \rightarrow 1$ limit.
For $n \neq 1$, there is a conical singularity at the origin of the Riemann surface and we have presented a detailed
analysis of the structure of the ``boundary" excitations of the $O(N)$ CFT at this singularity. 
(A closely related CFT with vortex boundary conditions was studied in Ref.~\onlinecite{VBSVortex} with a very different physical
motivation.)
In particular, we showed that in the context of $\epsilon = 4-D$ expansion,
the RG flow of the boundary coupling $c$ in Eq.~(\ref{boundary}) was the key to a determination of the
entanglement entropy. The RG flow of $c$ had two possible structures shown in Figs.~\ref{Figbetac} c) and d).
For $n$ greater than a critical $n_c$, we had flow in the infrared to $c=-\infty$ as in Fig.~\ref{Figbetac} d).
In contrast for $n<n_c$, we had three possible fixed points, and the $n \rightarrow 1$ limit
was controlled by the non-zero fixed point $c=c_r^+$, at which all strong hyperscaling assumptions
were obeyed. All our computations in the $\epsilon$ and $1/N$ expansions were consistent with this RG flow and fixed-point structure. 
One crucial consequence of the boundary perturbation and the subtle limit $n \to 1$ is that the finite size and correlation length corrections to the entanglement entropy are different at the Wilson-Fisher and Gaussian fixed points already at leading order in $\epsilon$ expansion. 

In this paper we have considered a geometry with a smooth, straight boundary between regions $A$ and $B$. One possible extension of our work is to consider boundaries with sharp corners. In such geometries, it is expected that the entanglement entropy will contain a universal logarithmically divergent term.\cite{Casini, Fursaev, Fradkin} Moreover, we have only studied the correlation length correction to the entanglement entropy in the symmetry unbroken region $t > 0$. It would be interesting to extend our treatment to the symmetry broken phase $t < 0$. 

While our paper was being completed, we learned of the numerical study of entanglement entropy
in the $d=2$ quantum Ising model in Ref.~\onlinecite{vidal}. At the quantum critical point the authors of Ref.~\onlinecite{vidal} find evidence for a finite size correction $\gamma$ as in Eq.~(\ref{gammadef}) in the case when the boundary between regions $A$ and $B$ is smooth. We note that the geometry studied in Ref.~\onlinecite{vidal} is an $L\times L$ torus divided into two equal cylinders rather than the infinite cylinder cut in half that we have considered here. Thus, the two results cannot be compared directly. Nevertheless, the value of $\gamma$ in Ref.~\onlinecite{vidal} is found to be positive, as in our conjecture in Eq.~(\ref{gammaloge}) for the case of periodic boundary conditions along the cylinder. 

\acknowledgements{We thank E. Fradkin, J. Zaanen and Y. Ran for useful discussions.
The research was supported by the NSF under grant DMR-0757145, by the FQXi
foundation, and by a AFOSR MURI grant.
}

\end{document}